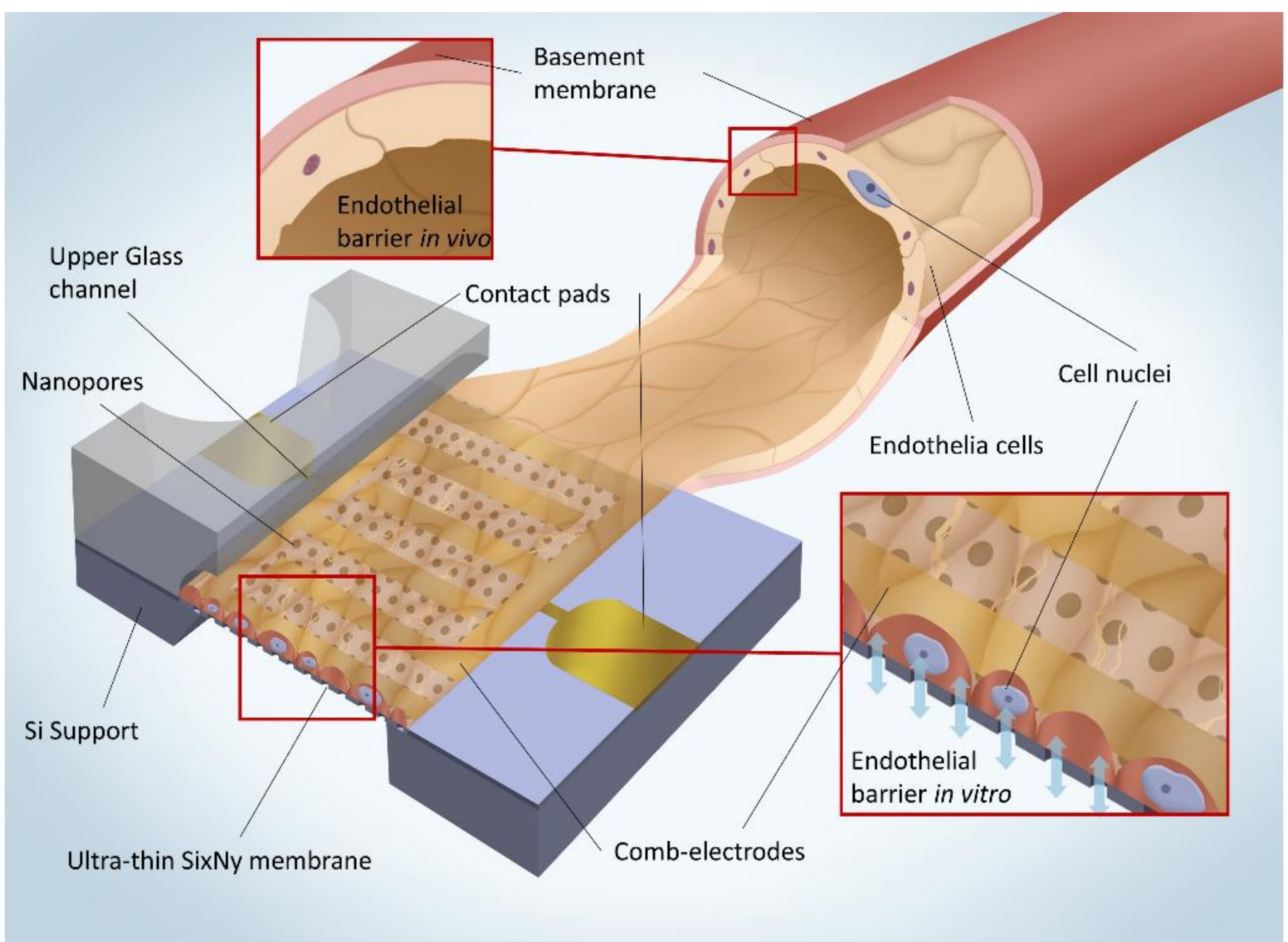

**Graphical Abstract**

**Smart membrane – high-content in-situ monitoring barrier -on-chip with artificial neural network**

*Bo Tang* [*]*, Victor Krajka, Mengxi Liu, Wei Zhao, Gazal Gökkus, Paul Lukowicz, Lili Zhu, Pu Chen, Stephan Reichl and Andreas Dietzel* [*]

Bo Tang, Victor Krajka, Wei Zhao, Gazal Gökkus, Andreas Dietzel

Institute of Microtechnology,

Technische Universität Brauschweig,

Brauschweig, 38124, Germany

Email: b.tang@tu-braunschweig.de, a.dietzel@tu-braunschweig.de

Bo Tang, Victor Krajka, Stephan Reichl, Andreas Dietzel

Center of Pharmaceutical Engineering (PVZ),

Technische Universität Braunschweig,

Braunschweig, 38106, Germany

Mengxi Liu, Paul Lukowicz

German Research Center for Artificial Intelligence (DFKI),

Kaiserslautern, 67663, Germany

Stephan Reichl

Institute of Pharmaceutical Technology and Biopharmaceutics,

Technische Universität Braunschweig,

Braunschweig, 38106, Germany

Lili Zhu, Pu Chen

Tissue Engineering and Organ Manufacturing (TEOM) Lab

Department of Biomedical Engineering

Wuhan University TaiKang Medical School (School of Basic Medical Sciences)

Wuhan 430071, China.

Conventional transepithelial electrical resistance (TEER) technique provides only a low-content analysis of cell-layer conditions, necessitating repeated microscopic assessments of morphology and cell-cell contacts outside the incubator for barrier-on-chip systems. This work presents a novel high-content TEER device in the form of a novel nanoporous membrane that facilitates continuous electrical measurement of cell-substrate impedance sensing (ECIS). The ultrathin (700 nm) membrane, composed of ultra-low-stress $Si_xN_y$, is monolithically integrated into wafer-level fabricated chips sealed with glass lids. Coplanar ECIS electrodes were connected to custom electronics to record impedance under sinusoidal excitation. Human umbilical vein endothelial cells (HUVECs) were seeded and continuously recorded impedance spectra were compared with bright-field and fluorescence microscopy, revealing distinct phases of monolayer formation. With one-dimensional convolutional neural network (Conv1d) and Kolmogorov-Arnold Network (KAN) trained with a small amount of Nyquist-diagrams, phases of (I) adherence, (II) outspreading, (III) confluence and (IV) barrier maturity with tight junction formation could be recognized with 95% confidence. As further proof of concept, reversible and irreversible barrier weakening using modulators PN159 and BAC was identified in this way. Our studies have demonstrated that an immediate and automatable non-invasive detection of *in-vitro* barrier dynamics within barrier-on-chip systems, eliminating the need for microscopy and endpoint staining. We expect this ECIS technique will find broad applications in organ-on-chip systems for in situ monitoring physiological or pathological states of tissue barrier.

## 1. Introduction

During the last decade, organ-on-a-chip technology has evolved tremendously. The most commonly used organ-on-chip architecture employs porous polymer membranes separating two microfluidic channels and is built using polydimethylsiloxane (PDMS), other polymers, or glass[1,2]. Such devices can support basic medical research into cellular behaviors, mimic human physiology and model human diseases.[3–5] In pharmaceutical research, organ-on-chip can be applied in drug screening and even reduce animal experiments. The development of cell barrier-on-chip for permeation studies has also made great progress,[6,7] including previous work of our group on vascular, ocular, and nasal barriers, and microfluidic chips enabling self-loading and endothelial polarization in pumpless flow.[8–11]

Barrier-on-chip platforms can provide a fast, non-invasive, effective, and extensively automatable tool that can be used to determine the integrity and confluence of cell monolayers, as R&D in the pharmaceutical industry requires the reliable generation of large amounts of reproducible data to allow predictions for (pre-) clinical studies. In addition to empirical judgment based on bright-field and fluorescence microscopy images, electrical, electrochemical, and photoelectrochemical methods have also been used on barrier-on-chip platforms for cell characterization. Transepithelial electrical resistance (TEER) measurements have been widely used as an effective tool for non-invasive and rapid monitoring of epithelial cell-barrier integrity and cell-cell tight-junction formation in *in-vitro* tissues, including ones cultured on chip.[12–14] Introducing on-chip TEER sensors has successfully eliminated uncertainties caused by operator influences during manual TEER measurements.[8,15] However, since TEER only measures the overall resistance (or impedance at a fixed frequency) through a cell-monolayer cultured on a membrane, a complementary assessment of cell morphology and the maturation state of cell-cell contacts by microscopy is still required.[13]

The light-addressable potentiometric sensor (LAPS) is a semiconductor-based platform for chemical and biological sensing that has made progress in the imaging of adherent cells.[16,17] LAPS enables fast imaging of single-cell surfaces using biased electrolyte-insulator-semiconductor structures in combination with fast scanning focused laser beam. Despite its great potential, this technique requires a continuous insulating layer ($SiO_2$ or $Si_xN_y$) between the cells to be imaged and the semiconductor,[17] which precludes its use in permeation studies where the cell barrier is grown on porous membranes. Other methods used for in-depth characterization of cell monolayer functionality and integrity include reflection impedance microscopy (RIM) and scanning electrochemical microscopy (SECM).[18,19] However, such methods require optical lens sets or scanning probes, making their integration in organ-on-chip systems difficult. Electrical impedance tomography (EIT) is another relatively new method with enormous potential in 3D imaging of cell spheroids and cell aggregates. In a typical EIT setting, small currents are injected into cell cultures using circular electrode arrays surrounding the sample, and a 3D map of conductivity is generated using complex mathematical algorithms.[20–23] EIT imaging is commonly used for 3D culture monitoring and has not yet been applied to cell suspensions or cellular monolayers.[21,23]

An alternative approach for characterizing cell monolayers is electrochemical impedance spectroscopy (EIS), in which sinusoidal voltage signals with varied frequencies are applied to the cell monolayer. With the help of coplanar electrodes, it is possible to obtain information about the electrical impedance of cells attached or in proximity to the electrode surface. Fitting data to create Nyquist diagrams using commercial programs or open-source algorithms results in numerical values that can be used in combination with other information about the state of the cell samples (e.g., microscopy) to create equivalent electrical circuit models for which good fits are obtained. However, interpreting EIS data remains challenging as more than one circuit model may lead to a good data fit, and the cultured tissue structure is complex and dynamic.[24,25] Electrical cell-substrate impedance spectroscopy (ECIS) is a variant of EIS technology that measures barrier integrity while providing information on cell adhesion, spreading, and growth. This is made possible by placing electrodes directly below and in contact with the cell surface. However, current studies are unable to provide a qualitative assessment of the dynamics of a forming cell monolayer,[26–28] which is significantly hindered by the overly complex biochemical reactions and unpredictable cellular metabolites in cell culture systems.[29,30] As a result, the equivalent circuits obtained are only some of the countless possibilities to represent the measured impedance values. Hence, there is no way to determine how well the equivalent circuits approximate the real system. For this reason, researchers make more or less simplistic assumptions in each case. Accordingly, automation of electrochemical interpretation, or an end-to-end solution, is yet an unsolved challenge for ECIS.[25] Nevertheless, ECIS has been successfully used with adherent cells for cell growth rate measurements,[31] with cell suspensions for cell concentration measurements,[23,32] with neuronal cells for monitoring cell adhesion, cell growth, and morphological changes during neuronal differentiation,[33] and even with lipid bilayers to measure the diameter of pores.[34]

Up to now, ECIS measurements have never been performed on barrier-forming tissue with coplanar microelectrodes directly integrated into ultrathin nanoporous membranes in organ-on-chip systems. In these, it is necessary to reproducibly record all stages of monolayer formation from initial cell attachment to the formation of a dense, mature monolayer with a uniform tight junction distribution, which until now has always required microscopy (e.g., end-point staining) in addition to TEER measurements. To eliminate the need for microscopic monitoring, we have developed a barrier-on-chip platform using Si as our chip material and nanoporous $Si_xN_y$ as membrane. Human umbilical vein endothelial cells (HUVECs) were employed as a model for barrier formation. Coplanar microelectrodes were used to monitor the electrical impedance of cells. The aim was to automatically detect different phases of monolayer formation and the establishment of a cellular barrier based on in-situ obtained electrical data alone, using a neural network-based method as an end-to-end solution.

## 2. Results

### 2.1. Barrier-on-chip system with electric ultra-thin nanoporous membrane

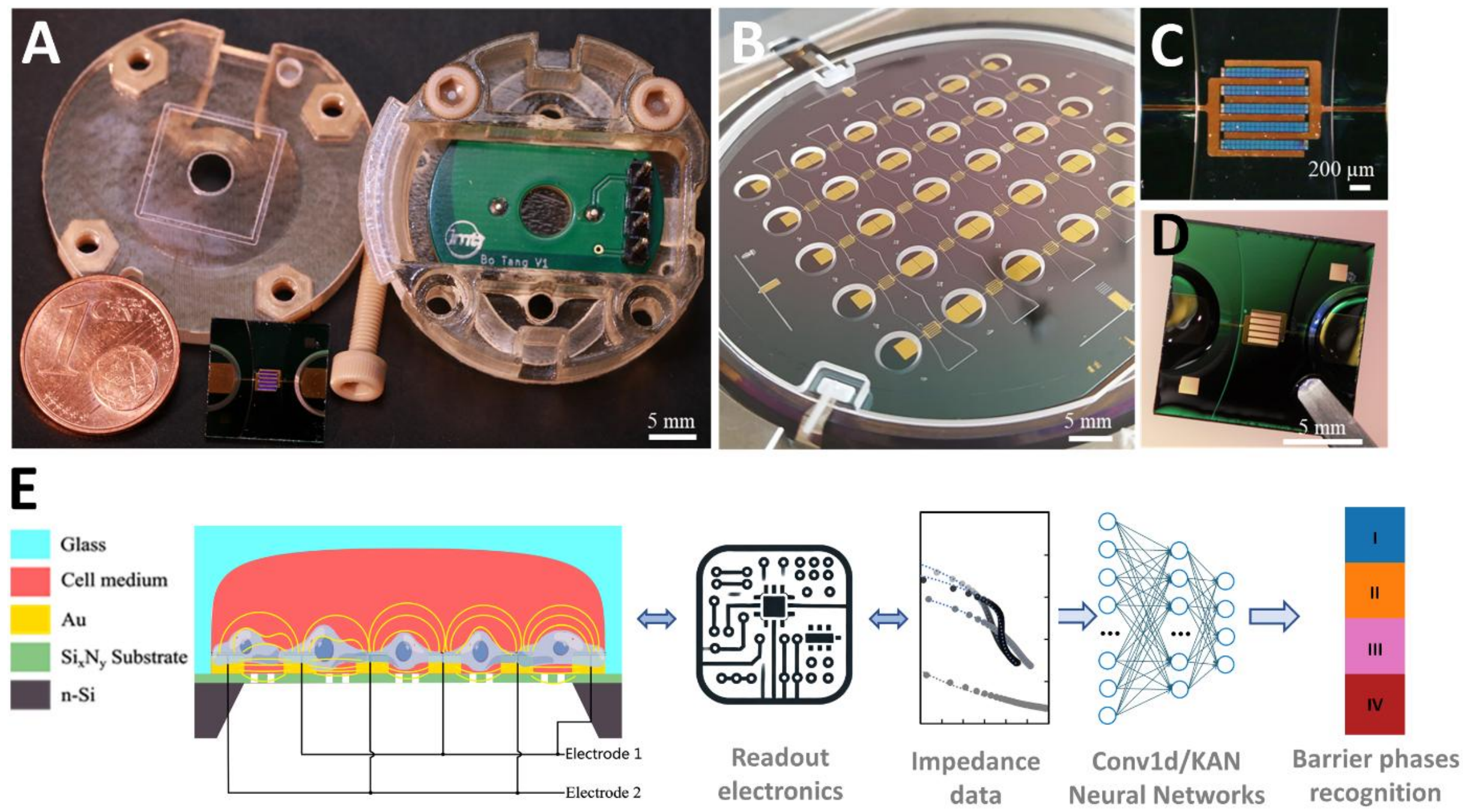

**Figure 1.** Microfluidic device for the detection of barrier formation and schematic illustrating of the electronic setup for impedance measurement on smart membranes and subsequent data processing. Excitation and readout electronics were connected to the comb electrodes onto which cells were directly cultured. (A) Photo of opened 3D printed pumpless fluidic supply system with integrated PCB for impedance reading. (B) Photo of structured glass wafer aligned with nano-fabricated $Si_xN_y$ wafer in anodic bonding chuck. (C) Micrograph of the interdigitated electrodes on the nanopore array (bluish shimmer). (D) Finished electric membrane chip. (E) Cross-sectional view of the schematic within the chip flow channel, where the distribution of the electric field generated by electrode 1 and electrode 2 varies depending on the density and morphology of the cells. The signal path leads from the PCB via the readout electronics to processing with artificial neural networks.

Our barrier-on-chip system consisted of several components, including the 3D-printed chip holder with medium reservoirs, microfabricated chip, and readout PCB (**Figure 1A**). The PCB had four connectors; two were used for signal readout and stimulation. The two parts of the holder were sealed by screwing the two components together with (polyetheretherketone) PEEK screws and grouting an O-ring. **Figure 1B** shows structured glass and $Si_xN_y$ wafers aligned and placed together in an anodic bonding chuck (EVG GmbH). A closer view of the porous membrane placed between the comb electrodes is shown in **Figure 1C**. With the exception of the electrodes, the cell culture area of the ultra-thin membrane was highly transparent due to the small membrane thickness (700 nm). In earlier work, the membrane pore size of 500 nm was found sufficient to allow the exchange of fluids, nutrients, or even drugs while simultaneously preventing the migration of cells through the membrane.[9] A single chip of 1 cm × 1 cm size, as obtained after wafer dicing, can be seen in **Figure 1D**. The smart membrane region consisting of the porous array and the electrodes was 3 mm × 3 mm in size.

The smart membrane represents a two-electrode impedance measurement system encapsulated in a microfluidic channel (see **Figure 1C, E**). For ECIS measurements, a 5 mV excitation amplitude was used to avoid undesired electrochemical reactions. An impedance readout circuit was configured based on a microcontroller (STM 32) (see **Figure 1F**). An analog-front-end impedance converter was directly connected to the two electrodes. A 12-bit digital-analog converter generated the excitation signal. The resulting current signal was measured by a current-voltage converter (TIA) followed by a 16-bit analog-digital converter so that the microcontroller could process the digital impedance data using Fast Fourier Transformation. Impedance data were then fed into Conv1d and KAN neural networks for training the model or recognition of barrier formation phases.

## 2.2. Tissue formation by bioimpedance and microscopic characterization for AI training

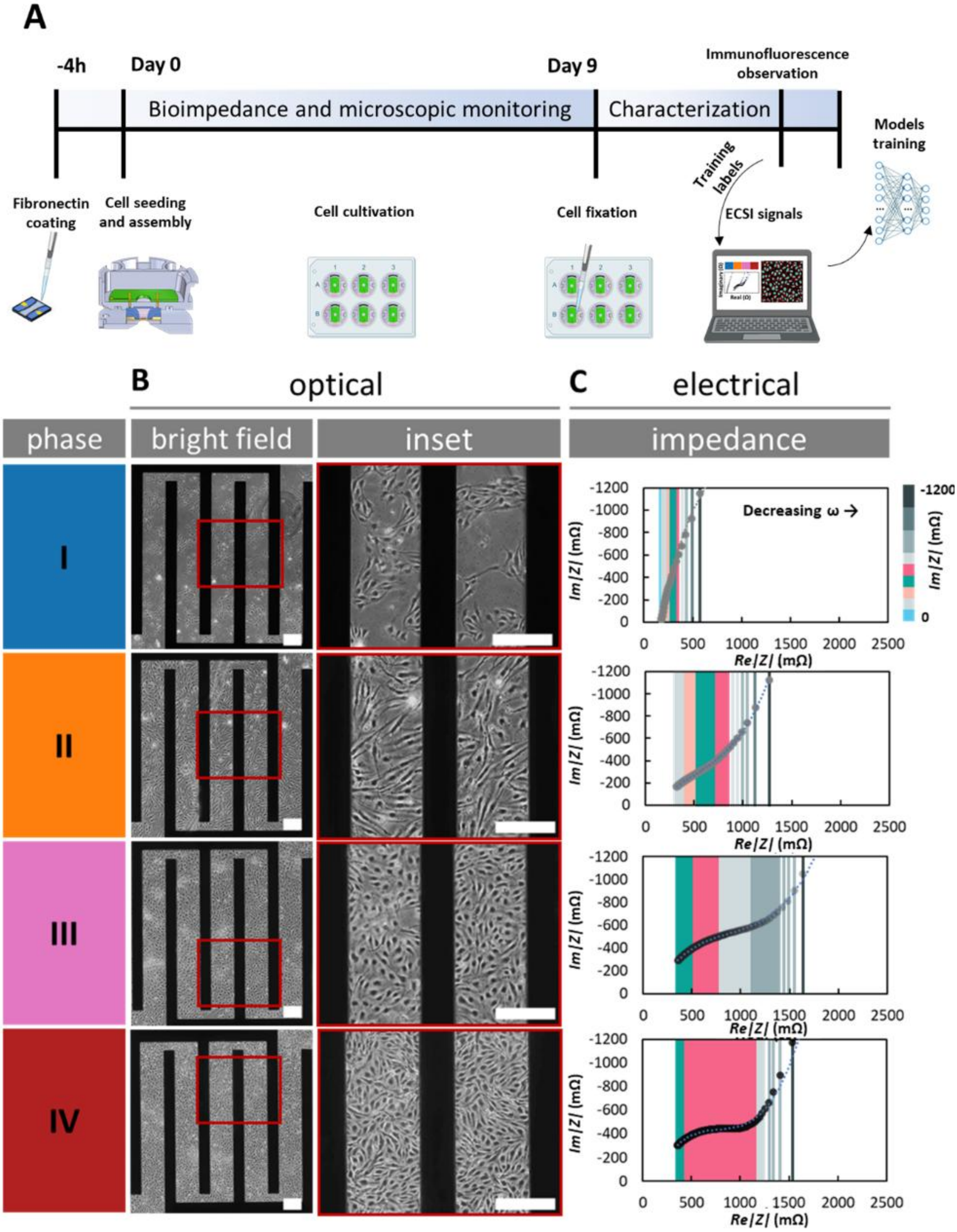

**Figure 2.** Electrical and optical monitoring of dynamic endothelial barrier formation. (A) Workflow of cell cultivation, bioimpedance spectra collection, microscopic observation and artificial neural network training. (B) Bright-field microscopy images of HUVECs corresponding to each phase of barrier formation. Phase I (blue): adherent cells are covering only a fraction of the membrane one hour after inoculation; phase II (yellow): cells attempt to spread out but have not formed a confluent layer yet; phase III (pink): cells continue to divide

to form a confluent monolayer; phase IV (red): formation of a very dense monolayer. Cell density and morphology can be tracked in the insets: As cell density increased, cell morphology changed from elongated, spindle-shaped to cobblestone-like. Scale bars represent 200 µm. (C) Nyquist diagrams show the electrical assessment of barrier formation. The frequency spectrum ranged from 1000 Hz to 200 kHz at equal intervals. With increasing cell coverage of the membrane, the Nyquist graph changed from an initially almost straight line to a characteristic semi-circle in the intermediate frequency range. The graphs are color-coded in the background based on the $Im|Z|$ values (between 0 and -600 Ω, the color changes at 100 Ω intervals, between -600 and -1200 at 200 Ω intervals).

Human endothelial cells were chosen as a biological model, as they are the principal regulators of vascular permeability *in vivo*.[35] Due to the high cost of the small batch wafer processing and for later cross-system validation, glass-bottum chips with the same electrode design were partially used instead of $Si_xN_y$ nanopore chips during data collection. After fibronection coating, the cells were seeded and the devices were mounted in a 3D-printed holder with circuit board and inserted into a standard 6-well plate. The cells became adherent within one hour of incubation. ECIS measurements were performed three times a day (8 am, 12 pm, 4 pm) using a sinusoidal stimulation peak-to-peak voltage of 5 mV in the frequency range of 1000 Hz to 200 kHz. Each measurement was conducted three times. To obtain sufficient data for neural network training, 18 glass-bottom chips and six nanopore chips were used. Cells were cultured for up to 9 days and each electrical measurement was accompanied by bright field images using an inverted microscope (IX50, Olympus) to determine cell density and morphology. Representative chips were selected for subsequent cell fixation and immunofluorescence staining, Once the phases of barrier formation had been determined by microscopic images of cell morphology and fluorescence staining, the ECIS data could be used to train artificial neural network models (see **Figure 2A**). As shown in **Figure 2B**, the formation of a dense monolayer proceeded in four color-coded phases. Initially, the cells were spreading (phase I in blue), and through continuous cell proliferation, the cell density increased but cells remain with elongated morphology (phase II in yellow). When a confluent monolayer (phase III in pink) was formed, the cell morphology appeared more round. Eventually, the monolayer reached its maximum cell density (phase IV in red). In parallel, the monolayer formation was monitored via ECIS measurements. The obtained impedance values can be represented in complex form, consisting of a real $Re|Z|$ and an imaginary part $Im|Z|$. As is common practice [36], our ECIS data are presented as Nyquist diagrams (**Figure 2C**), where each data point corresponds to the impedance at one frequency. The leftmost data point corresponds to the impedance measured at the highest frequency (200 kHz), and the rightmost data point corresponds to the lowest frequency (1000 Hz). To alleviate the comparison between the patterns of Nyquist diagrams and to make the changes more visible, the Nyquist diagrams are color-coded based on the imaginary values. Areas with a lower slope have a wider and areas with a higher slope a narrower color band (**Figure 2C**). In phase I, only a small number of cells (occupied membrane area ≤ 30%) are present on the membrane surface. This results in an almost complete absence of a semi-circle in the high-frequency region. The value of the intersection with the x-axis reflects the electrical resistance of the cell culture medium in the system. In phase II, the cell density increases, the membrane gradually becomes more covered (occupied membrane area > 30% and <99%), and a semi-circle appears. In phase III, a complete cellular monolayer is formed (occupied membrane area ≥99%) and the Nyquist graph assumes a typical Randles circuit pattern [36] in the high to intermediate frequency range, indicating that resistive and capacitive elements are connected in parallel. In phase IV (occupied membrane area ≥99%, but with higher cell density than phase III), the influence of the resistive and capacitive elements connected in parallel becomes even stronger, indicating that tight junctions are formed and that the electrical path through the cells carries more of the current.

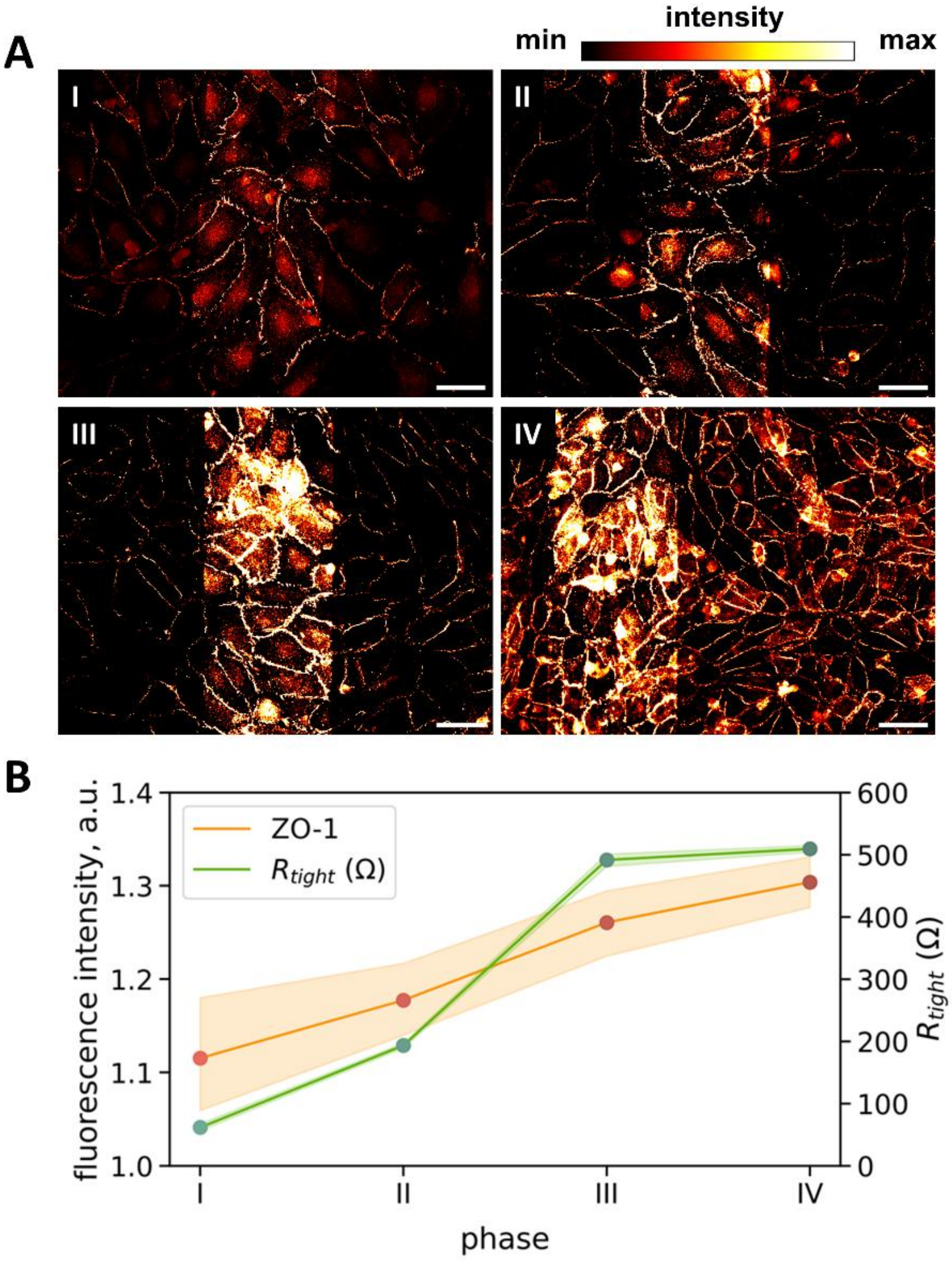

**Figure 3.** Emerging endothelial barrier showing in fluorescence micrographs and electrical measurements. (a) Images with immunofluorescence staining against tight junction-associated protein ZO-1 (intensity shown in pseudocolor red-hot). The intensity at cell-cell contact sites of ZO-1 increases with longer cultivation time. The time points correspond to 12h (phase I), 84h (phase II), 108h (phase III), and 132h (phase IV) after inoculation of the cells. Scale bars represent 25 µm. (b) Measured ZO-1 fluorescence intensities for the four phases of monolayer formation plotted together with tight junction resistance $\boldsymbol{R_{tight}}$. Increased reflectivity in the metallized areas was compensated based on the ratio of cell outline to cytoplasmic intensity normalized to the starting point (**Supporting Information Note 1**). The error band represents a 95% confidence interval (N = 3).

With impedance analysis software (ZView from Scribner Associates, Inc), the Nyquist diagrams were analyzed assuming an equivalent circuit model that describes the nanoporous membrane chip system with cell coverage. Equivalent circuit analysis is a fundamental method[36] in electrochemical analysis. The assumed equivalent circuit elements and the values obtained for them are described in detail in **Supporting Information Note 2**. From phases I to IV the paracellular tight junction resistance $\boldsymbol{R_{tight}}$ shows a monotone increase with the highest average value of 510 Ω in phase IV. This was proven to correlate with the formation of dense, tight junction contacts as observed by immunofluorescence staining against the tight junction-associated protein ZO-1, which also continuously increased from phase I to phase IV (**Figure 3a,b**).

### 2.3. Impedance spectrum chronogram

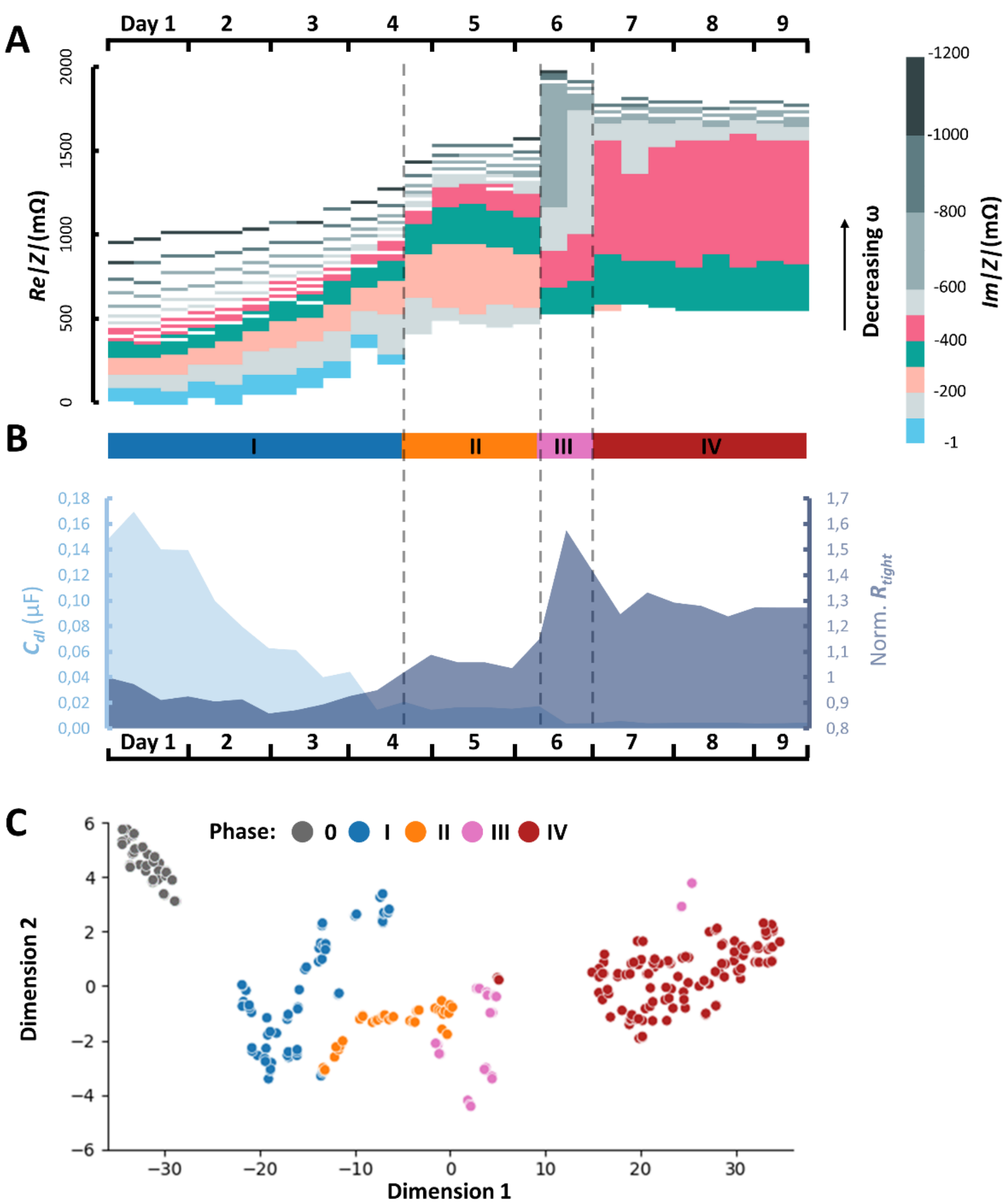

**Figure 4.** ECIS data was obtained with HUVECs that were cultured for one week in less costly to produce chips with microelectrodes on the bulk glass bottom. (A) Nyquist chronogram based on the *Im*|*Z*| values (same color coding as in previous **Figure 2**). (b) Course of double layer capacitance $C_{dl}$ and ***Norm.*** $R_{tight}$ (which is $R_{tight}$ normalized to its start value). (c) Scatter matrix for 2-dimensional t-SNE analysis of the Nyquist data sets from all barrier formation. Phase 0 represents non-adherent cells immediately after inoculation.

**Figure 4** shows the evaluation of impedance data obtained by culturing cells on 24 less costly glass bottom chips for one week. To visualize the impedance changes over time, we have replaced the individual Nyquist diagrams by color-coded imaginary impedance values. With this reduction of information depth, it becomes easier to follow the evolution of the barrier over time (**Figure 4A**). In addition, the course of the double layer capacitance $C_{dl}$ forming on the electrode surfaces and $R_{tight}$ (both obtained by fitting the Nyquist graphs as described in detail in **Supporting Information Note 2)** during monolayer formation is shown (**Figure 4B**). Within the first 2 days (phase I), the Nyquist-chronogram gradually shifts, and $R_{tight}$ slowly increases. In this phase, the cells are spreading over the membrane (and electrode) area without considerable formation of cell-cell contacts. With higher cell densities, the double layer formation on the electrode reduces, leading to a reduction of $C_{dl}$. The second phase was reached when the cells have spread semi-confluently. During this period (approximately one day), the Nyquist chronogram and $C_{dl}$ remained rather stable. The fastest dynamics were observed in phase III (<12 hours). Here, $C_{dl}$ decreased sharply towards 0 while $R_{tight}$ increased strongly. This indicates the formation of a confluent monolayer impeding charge transfer. Finally, in phase IV, a state of equilibrium was reached, where no pronounced changes in the Nyquist-chronogram could be observed. In this stable phase (until the end of the experiment), a high $R_{tight}$ plateau was reached. This coincided with the highest cell density and matured cell-cell contact formation. In summary, the progressive endothelial barrier formation can be divided into four phases already based on the color-coded Nyquist chronogram pattern and the course of $C_{dl}$ and $R_{tight}$. In the following, we will investigate whether a neural network can recognize the phases independently of human interpretation, thus enabling automation.

### 2.4. Data clustering and t-distributed stochastic neighbor embedding analysis (t-SNE)

Before training a neural network to detect the four phases, the impedance data set was dimensionally reduced using t-SNE to verify whether the four growth phases could be identified without using the data acquisition time. The data also covered the inoculation phase 0 in which the cells had not yet attached to the sensor surface. The algorithm first calculated the unscaled similarity between each point by normal distribution, then randomly projected these points onto a two-dimensional plane. Then, the unscaled similarity between all the points and the points on the 2D plane was computed via the t-distribution function and by shifting the points on the 2D plane until the t-distribution similarity of each point on the 2D plane was the same as the normal distribution similarity in the high-dimensional space. This dimensionality reduction is intended to help with the clustering of high-dimensional data. The scatter matrix obtained from the two-dimensional t-SNE analysis is shown in **Figure 4C**, which reveals that in the plane of dimension 1 and dimension 2 the data exhibit four distinct clusters that could be directly associated with the four phases of barrier formation. Phases I and IV, represented by the blue and red colors, are clearly separated, while phases II (orange) and III (pink) slightly overlap with the adjacent clusters. The cluster 0 (gray) represents measurements taken directly after inoculating the cells into the chip. While it remained difficult to distinguish between cells in suspension and adherent cells through the subjective viewing of Nyquist diagrams, this stage became clearly recognizable in t-SNE. The fact that the collected impedance data sets could be clustered by dimension reduction permitted the subsequent analysis with a neural network.

### 2.5. Neural network recognition of tissue evolution phases

The impedance data collected from the glass bottom chips were divided into two parts: the first for training and the second for pre-validation. The pre-validated neural network should then be used to predict the state of the barrier also in the nanoporous membrane chips. The confusion matrixes shown in **Figure 5** demonstrate the classification results (averaged over 20 repetitions) obtained with validation datasets from glass bottom chips (**Figure 5A, D**) and from ultra-thin nanoporous membrane chips (**Figure 5B, E**). The recall score of Conv1d (determined as described in the methods section) was in all phases greater than 79% for the glass-bottom chip and 80% for the nanoporous membrane chips (**Figure 5A, B**). For the glass-bottom chip, the model recognized phases I (100%) and IV (99%) reliably (**Figure 5A**). In contrast, phases II and III were robustly detected (79-82%), with ~10% misclassifications to the adjacent phases. This is in line with the previous t-SNE analysis, where the clusters of phases II and III displayed a certain overlap. In contrast, the recognitions for the mature barrier stages (phases III and IV) were very accurate (94% and 93%) for the nanoporous membrane chips, while the correct classification for immature barrier stages was a little lower (80% and 84%) (**Figure 5B**). The recognitions obtained with a Kolmogorov-Arnold network (KAN) were even better. KAN model could better recognize the phases II and III (84-89%), even though KAN was trained only with 30% of glass bottom chips and Conv1d with 80%. The maturation of the barrier depends on many factors, namely cell number, cell area, cell-cell contacts, etc. Each of these factors changes at different rates, and therefore the duration of the phases varied. More data sets were acquired for longer phases and fewer for short phases. Since this imbalance in the number of data sets was not compensated for, the simple recall calculation cannot fully reflect the model's performance. For this purpose, the precision (1 - false positive rate), the recall (1 - false negative rate), and the macro F1-score (harmonic mean of precision and recall) were calculated for all phases of barrier formation, and the data distribution was compared for both chip types (**Figure 5C, F**). Generally, the overall precision (~95%), recall (~94%), and macro F1 score (~94%) were higher for the glass bottom chips compared to the nanoporous membrane chips (~89%, ~87%, ~87%) for Conv1d (**Figure 5E**). However, the KAN model recognized the phases in nanoporous membrane chips as good as in glass-bottom chips (**Figure 5**F**)**, indicating that the KAN model exhibits better generalization. Overall, the neural networks showed that the models were able to reliably recognize the most mature stages of the endothelial barrier, while the more dynamic intermediate stages were less accurately detected. Since the neural network was trained with ECIS data from the glass bottom chip, a slightly less accurate reorganization of barrier formation phases in the nanoporous membrane chips was to be expected. Since a dense, mature barrier is the starting point for most drug permeation studies, it was most important for us to identify this phase reliably.

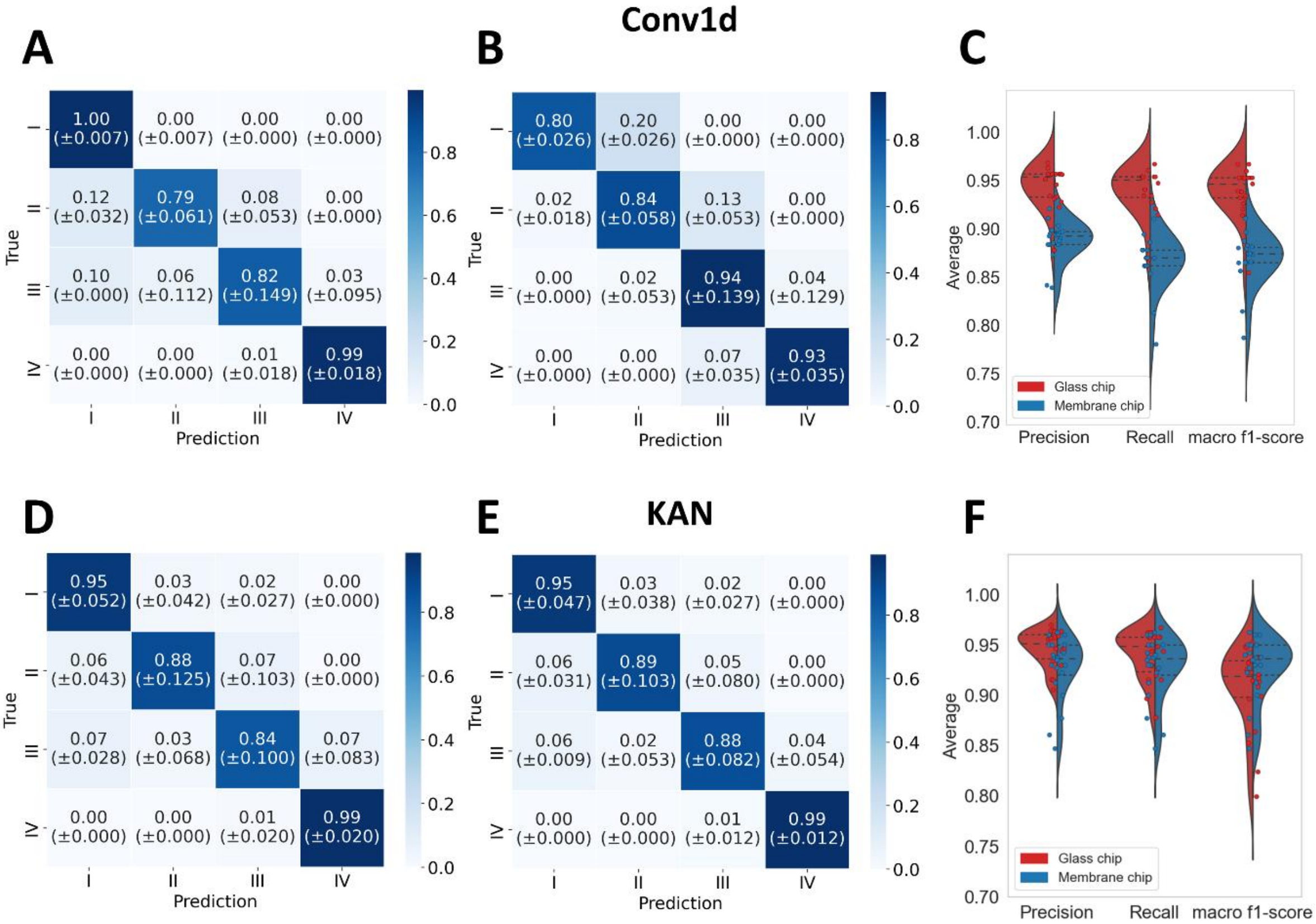

**Figure 5.** Performance reports of Conv1d and KAN neural networks. (A, D) Confusion matrixes for self-evaluation of the trained models with glass bottom chips. (B, E) Confusion matrixes for cross-system validation with nanoporous membrane chips. (C, F) Violin plots of averaged metrics of the test results from glass bottom and nanoporous membrane chips.

To further discuss the performance of the model for different phases, we performed statistics on all the performance metrics for four defined cultivation phases. The results are shown in **Figure S3 (A, B).**

### 2.6. Recognition of phases in response to barrier modulators

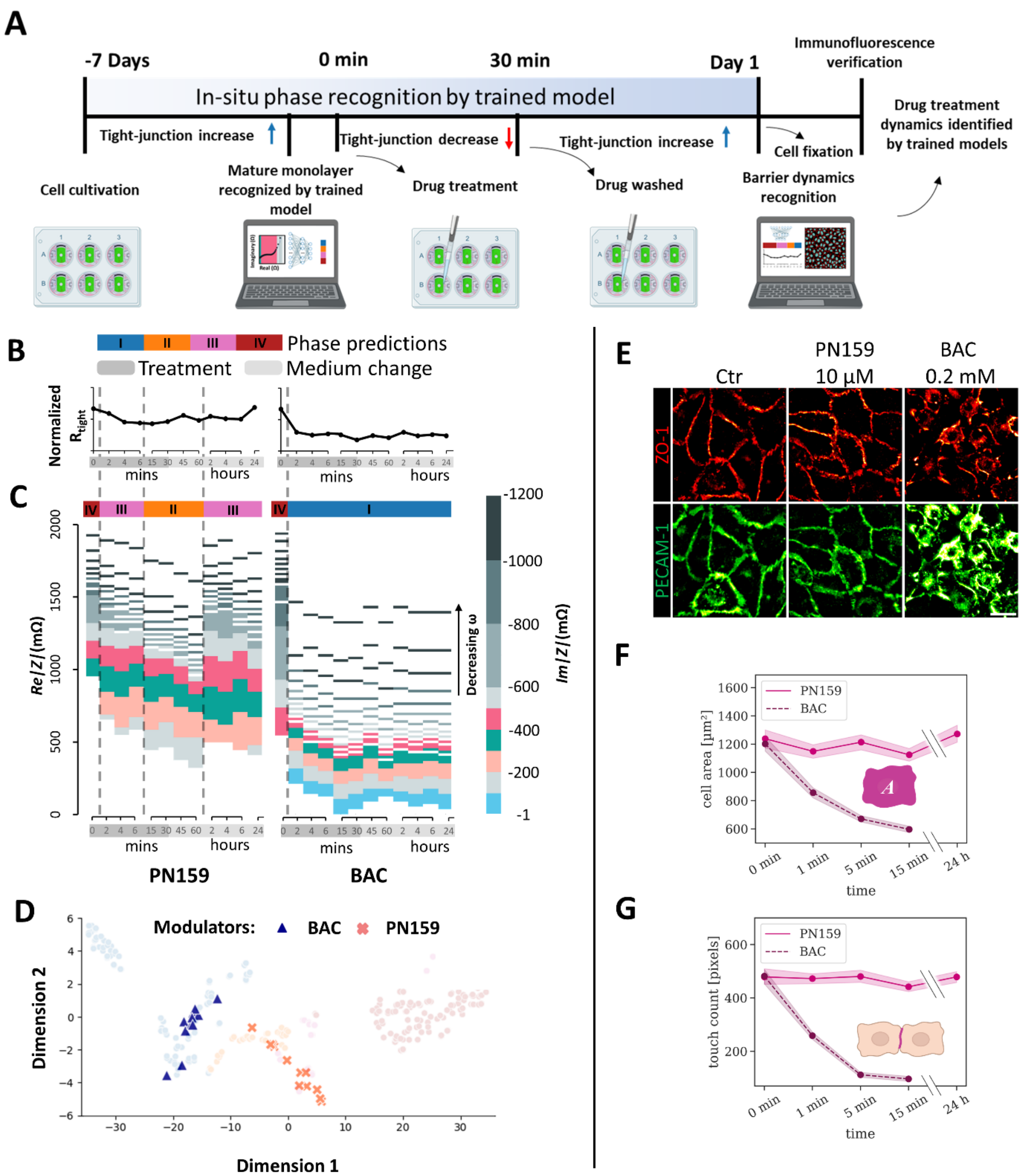

**Figure 6.** Influence of modulator treatments (PN159 10 mM and BAC 0.02%) on recognized phases of barrier formation. (A) Workflow of drug treatment and phase recognition by trained models. (B) Dynamic change of the ***Norm.*** $\boldsymbol{R_{tight}}$ for modulator treatments. (C) Nyquist chronograms based on the *Im*|*Z*| values show substance-specific monolayer integrity during and after treatment. (D) Scatter plot for clustering after drug treatment superimposed on data from untreated cultivations (BAC: dark blue triangles, PN159: pink crosses). (E) Micrographs depicting immunofluorescence staining of cell-cell junction-associated markers ZO-1 (red) and PECAM-1 (green) after 15 min of drug treatment. Scale bar represents 10 µm. (F)

Quantification of cell area during and after drug application. (G) Quantification of cell-cell contact amount during and after drug treatment.

To investigate whether the trained neuronal network can also recognize the influence of barrier modulators, we treated the cells either with a peptide PN159 or with the cationic surfactant BAC (benzalkonium chloride). PN159 is known to be a tight-junction specific modulator, while BAC is used to permeabilize phospholipid bilayers [37–39]. After achieving a mature monolayer (verified optically and electrically), drug application was initiated by flushing the channel with 100 µl of drug-containing medium and measuring the change in impedance over time (2, 4, 6, 16, 30 minutes after drug application). After 30 minutes, the cell culture channel was washed again with 100 µl of medium and the measurement series was continued (0.75, 1, 2, 4, 6, 22, 24 hours after drug application). The trained artifical neural network models were used during the measurement to recognize the barrier formation phase which was verified by immunofluorescence staining afterwards (see **Figure 6A**). To test different reaction rates, two concentrations were used for each compound (PN159: 1 µM, 10 µM; BAC: 0.02%, 0.1%), and impedance spectra were recorded within the first 30 minutes after drug application (**Figure 6,** measurements with all concentrations see **Figure S4**). In general, the decrease in ***Norm.*** $\boldsymbol{R_{tight}}$ was greater at higher concentrations, with a drastic decline within 2 minutes for BAC (**Figure 6a and Figure S4a**). After 30 minutes, the substances were washed out, and impedance spectra were measured over 24 hours. It was found that PN159-treated cells recovered in a concentration-dependent manner in contrast to BAC treatment, indicating that the barrier modulation was reversible (**Figure 6b**, **Figure S4**). The t-SNE analysis of the impedance data after treatment revealed a clear grouping by drug: the data points from PN159-treated cells clustered between phase IV to phase II (mature-confluent to semi-confluent), while the data points from BAC-treated cells showed a clear grouping in phase I (isolated adherent cells) (**Figure 6c**). Interpreting the impedance spectra by the neural network revealed a concentration-dependent course of barrier modulation (**Figure S4b)**. In general, barrier integrity changed from phase IV (mature barrier) to phase II (semi-confluent) after treatment with PN159 and returned to phase III (confluent) after a wash step. At the lower concentration, the transition to phase II took longer, and the residence time in phase II was shorter compared to the higher concentration. Similarly, the higher concentration of BAC resulted in a faster phase change from phase IV (mature barrier) to phase I (single cells) compared to the lower concentration. However, once phase I was reached, it remained there until the end of the recordings, showing an irreversible course.

To evaluate whether the phase recognition could be confirmed at the cellular level, individual cell morphology in the monolayer, cell-cell contact intensities (as an indicator of tight junction strength), and cell-cell contact length were assessed by immunofluorescence staining (**Figure 6**d). In contrast to monolayer formation, in which the relative intensities of the cell junction markers ZO-1 and PECAM-1 increased with progressive maturation, no substantial (PN159) to minor (BAC) changes could be measured after drug application (**Figure S5a, b**). However, the strong cellular effect of BAC could be quantified on a morphological level (e.g., cell shrinkage, fewer cell-cell contact length), while no significant changes were observed after PN159 administration **Figure 6e, f**). Overall, the assessment of monolayer integrity-based neural network interpretation of impedance data appears to be sensitive enough to detect even small changes in the cell layer, whereas $\boldsymbol{R_{tight}}$ or endpoint analysis using standard immunofluorescence staining does not appear to be sensitive enough.

## 3. Conclusions and discussions

A barrier-on-chip platform with impedance-sensing microelectrodes deposited onto a $Si_xN_y$ nanoporous membrane was successfully developed. The chip was fabricated using wafer-level processes, as are available in typical MEMS foundries, where scale-up to mass production could be realized. A measurement setup has been developed to record ECIS data over an extended period. With a 3D-printed holder and a readout board, the chips were kept in 6-well plates under standard cell culture conditions. On this platform, endothelial barrier formation was monitored optically via the cell coverage on the sensor surface and via immunofluorescence staining of the tight junction-associated marker ZO-1 and endothelia-specific cell junction marker PECAM-1. It was found that with rising cell density the fluorescence intensity of ZO-1 and PECAM-1 increased. This is consistent with previous experiments on the dynamic formation of tight junctions at different endothelial cell densities [40]. Accordingly, the electrical evaluation of the maturing barrier with the in-situ ECIS sensor showed a decrease in $\boldsymbol{C_{dl}}$ and an increase in $\boldsymbol{R_{tight}}$, while the Nyquist diagram changed towards a typical semi-circle. This is particularly important for distinguishing between the third and fourth phase of monolayer formation, where the Nyquist plots exhibit a completely different pattern in the mid-frequency region. Typically, the time course of the high-content impedance measurements is displayed by superimposing the curves in 2D or 3D plots.[41,42] We found that color coding can help to recognize the four phases of barrier formation more intuitively. This representation is similar to the appearance of a spectrogram, which is often used with seismometer signals to distinguish and characterize different types of earthquakes.[43] A recognition of the monolayer evolution phases from the high-content impedance data independent of the subjective observer was successfully demonstrated based on Conv1d and KAN neural network models, trained with data from glass bottom chips. The cross-system validation for nanoporous membrane chips was successful and confirms the generalizability of the model and the stability of the sensor, the significant differences between the patterns of the treated cell monolayers were clearly visible. This smart end-to-end solution has shown that the retrograde effects induced by barrier modulators can also be very clearly recognized in an automated manner and reversible and non-reversible effects can be distinguished.

With the demonstrated evaluation of high-content data by neural networks, we enable an end-to-end solution that automatically recognizes the maturity states of the biological barrier on the chip without microscopic analysis and can help to decide when to start a permeation test. This is of great importance for reliable high-throughput testing with barrier-on-chip systems in the pharmaceutical research and development.

## 4. Materials and methods

### 4.1. Chip fabrication with wafer-level processing:

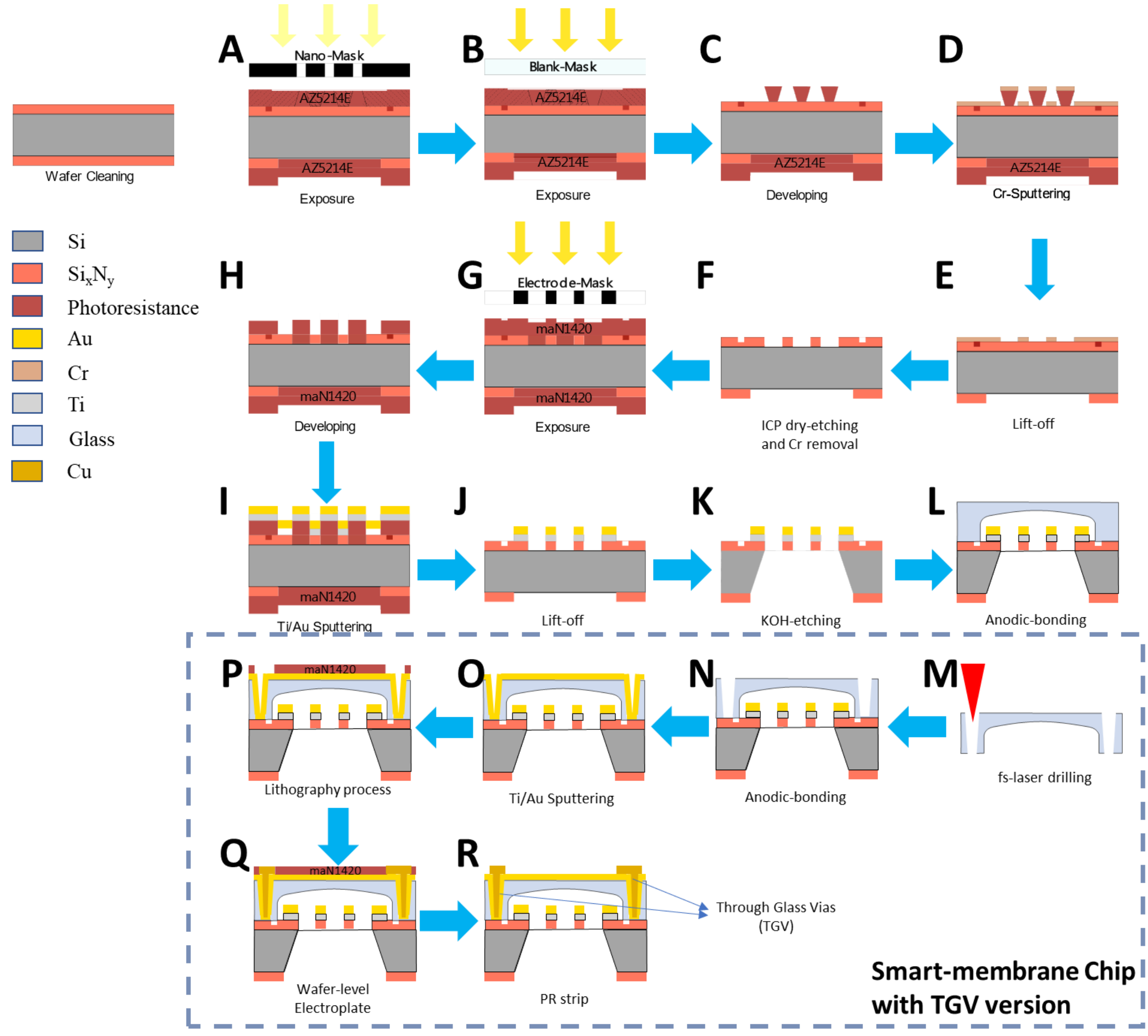

**Figure 7.** Schematic illustration of the fabrication of chips with Au/Ti electrodes on nanoporous ultra-thin $Si_xN_y$ membranes with scalable wafer-level processing. (a, b, c) Two step lithography process using a nano-mask in combination with image-reversal photoresist for nanoporous array structuring. (d, e) Cr-sputtering as a metal mask and lift-off process. (f) Dry-etching process to obtain array of nanopores in the $Si_xN_y$ layer. (g, h, i, j) Lithography followed by Au/Ti sputtering and lift-off for integration of electrodes on nanoporous layer. (k) Wet etching for releasing the nanoporous membrane. l) Anodic-bonding the Si/$Si_xN_y$ chip compartment with structured glass compartment to enclose the smart membrane in the glass channel. (p) Photolithography is performed to define the TGV metallization and contact-pad patterns. (q) Wafer-level Au electroplating is carried out on the sputtered Ti/Au seed layer to metallize the TGVs and thicken the conductive interconnects. (r) The photoresist is stripped to complete the TGV-based vertical electrical interconnections.

The microfabrication of chips is illustrated in **Figure 7**. A customized nanopore mask was purchased (Topan GmbH, 500 nm opening diameters). Nanopores (500 nm) were generated in 700nm layers of $Si_xN_y$ deposited on 100 oriented silicon monocrystalline 4-inch wafer substrates (< 35 MPa ultralow stress $Si_xN_y$ on silicon obtained from Hahn-Schickard-Gesellschaft Villingen-Schwenningen) by photolithography using image reversal high-resolution photoresist AZ5214E (Merck Performance Materials GmbH) and inductively coupled plasma (ICP) dry-etching (Cobra 300 ICP Oxford Instruments with 30 sccm SF6 1000W ICP process). A layer of 200 nm gold was sputtered (Laborsystem LS 440 S, Ardenne Anlagentechnik GmbH, Dresden, Germany) and structured by lift-off lithography before the $Si_xN_y$ membranes were released by KOH etching. Microfluidic channels (200 µm deep) were etched in borofloat-glass wafers (700 µm Borofloat® substrates from Schott AG, Mainz, Germany) using HF-solution (40% HF concentration). Anodic bonding was used to join glass and silicon wafers. Twenty five individual chips were produced on each wafer. Each chip was only 1 cm x 1 cm in size. Chip singularisation was performed by wafer dicing (DAD 320 Dicing Saw from DISCO Corporation). For the less complex glass-bottom chips electrodes were realized with the same pattern and same process but on bare borofloat glass wafers (700 µm Borofloat® substrates from Schott AG, Mainz, Germany). Glass substrates with electrodes were encapsulated using structured glass covers identical to the ones for the nanoporous membrane chip. Finally, the two glass compartments of the glass-bottom chip were fixed together with double-sided adhesive tape (ARcare® 90106NB).

### 4.2. Organ-on-chip system

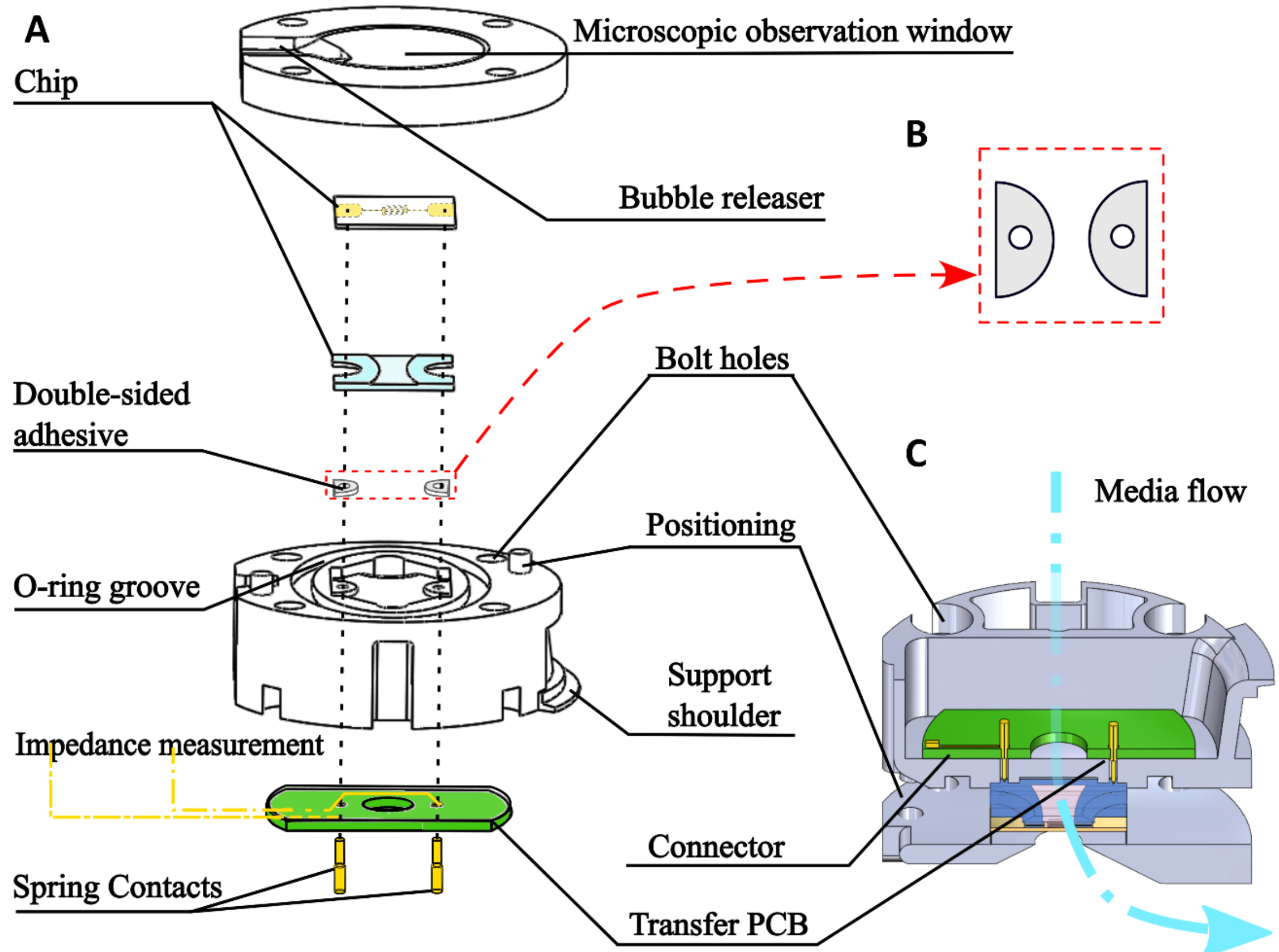

**Figure 8.** (A) Exploded view of the barrier-on-chip system. The system is held in a 3D-printed holder and is connected to a PCB through two spring contacts for impedance measurement. (B) Adhesive foil cut with a CO2 laser used to isolate the spring contacts. (C) Sectional view of the assembled system in the holder. Media flow is achieved through gravity driven pumping (blue arrow).

A holder for the chips with printed circuit board (PCB) was 3D printed (Keyence AGILISTA3200W) using AR-M2 transparent print material that was compatible with commercial 6-well plates. This setup allowed ECIS measurements in an incubator. In addition, the holder supports cells with medium and optionally allows gravity-driven flow through the chip. A compressed O-Ring sealed the fluidic stream. An observation window on the bottom of the holder allowed inspection of the chip membrane area using an inverted microscope. In addition, a bubble releaser is implemented in the area to release any air bubbles that may have accumulated (**Figure 8A**). A fluidic/electric isolation layer was applied in which biocompatible double-sided adhesives (ARcare® 90106NB, see **Figure 8B**) were compressed by the upper and lower part of the chip fluidic supply system with four polyether ether ketone (PEEK) screws (see **Figure 8A, B, C**). The spring contacts were soldered on a 4-layer transfer printed circuit board (PCB) in which the other two copper layers isolated the two signal lines. Using the spring contacts in the contact pad holes, a reliable electrical connection of the impedance readout device with the chip was established.

*4.2.1. Bioimpedance sensing system and readout devices:*

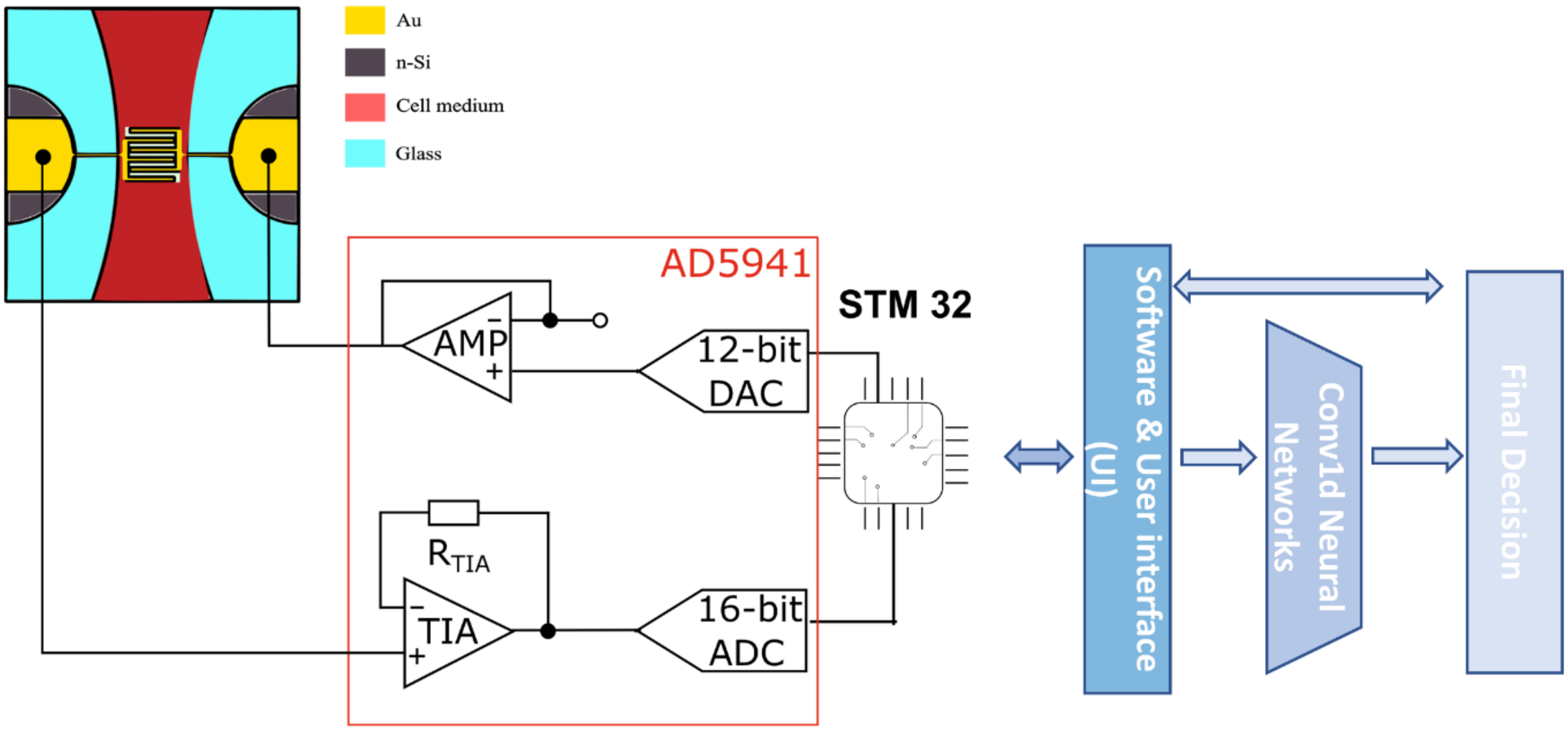

**Figure 9.** Schematic illustrating of the electronic setup for impedance measurement on smart membranes and subsequent data processing. Excitation and readout electronics were connected to the comb electrodes onto which cells were directly cultured. The smart membrane chip is connected to the readout electronics together with the data processing pathway.

The smart membrane represents a two-electrode impedance measurement system encapsulated in a microfluidic channel (see **Figure 9a**). For ECIS measurements, a 5 mV excitation amplitude was used to avoid undesired electrochemical reactions. An impedance readout circuit was configured based on a microcontroller (STM 32) (**Figure 9**). An analog-front-end impedance converter was directly connected to the two electrodes. A 12-bit digital-analog converter generated the excitation signal. The resulting current signal was measured by a current-voltage converter (TIA) followed by a 16-bit analog-digital converter so that the microcontroller could process the digital impedance data using Fast Fourier Transformation. Impedance data were then fed into Conv1d and KAN neural networks for training the model.

*4.2.2. Sensing Electric field distribution:*

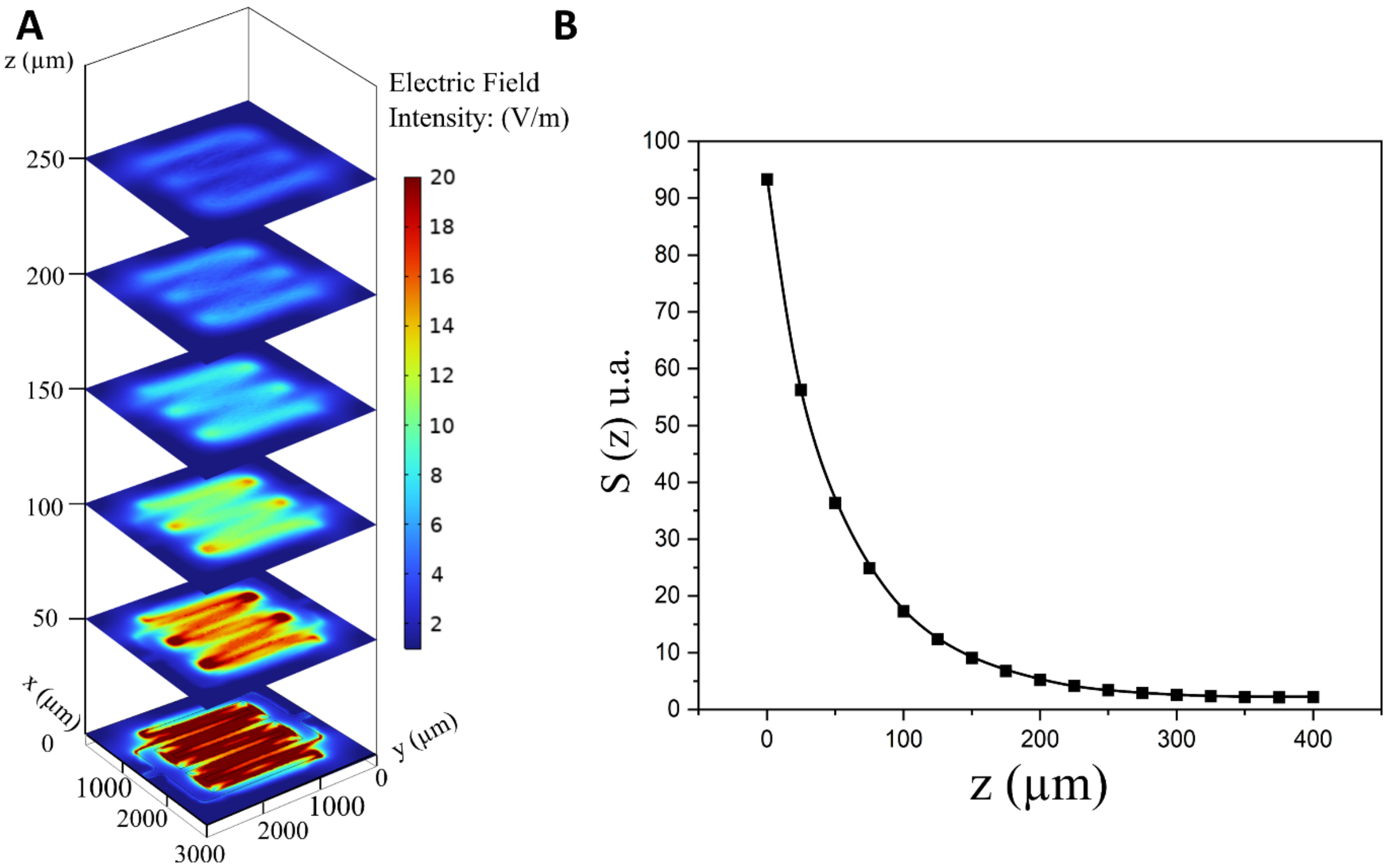

**Figure 10.** (A) Simulation of the electric field distribution at different heights starting from the electrode surface plotted with 50 μm intervals. (B) Sensitivity $S$ in dependence of height $z$ over the electrode surface.

The electric field intensity distribution directly affects the measurement sensitivity and range of the comb-electrodes sensor.[23,44] By simulating the electric field intensity, a sensitivity $\boldsymbol{S}$ can be calculated as

$$\boldsymbol{S(z)} = \int_{\mathbf{xy}} \frac{\mathbf{E(x,y,z)}}{\mathbf{V}}\, \boldsymbol{dxdy} \tag{1}$$

Where $\boldsymbol{E}$ is the electric field intensity when a voltage $\boldsymbol{V}$ is applied to the electrodes. In **Figure 10A**, a simulation of the electric field intensity in liquid environment starting from the electrode surface (0 on the z-axis) is shown. The sensitivity practically disappears at 200 μm distance from the electrodes. A sensitivity attenuation of less than 10% at 25 μm from the electrode suggests that the sensing electrodes are sufficiently penetrative to probe cellular monolayers with typical heights of 10-20 μm at an excitation voltage of 5 mV (**Figure 10B**). Not only cells cultured directly on the surface of the electrodes, but also cells residing on the cell culture membrane between the electrodes can be probed. However, measurements at the electrode surface and between the electrodes can have different contributions to the overall impedance because of the different orientation of field lines.

### 4.3. On chip cell cultivation protocol

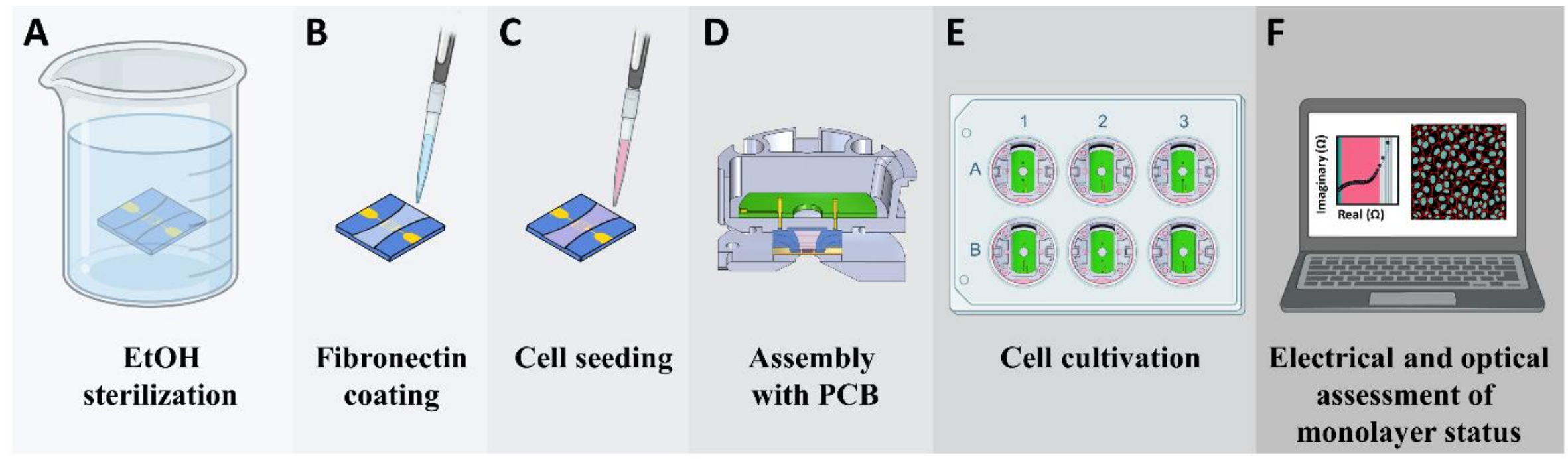

**Figure 11.** Schematic representation of the cell cultivation protocol. (A) Chips and 3D-printed holders were decontaminated by immersion in 70% ethanol for 20 minutes. (B) After the chips were dried under a biosafety cabinet, the chip channels were coated with fibronectin. (C) After aspiration of the coating solution, the channels were dried and a defined number of cells were injected into the channels. (D) After cell attachment (30 min), the chips were assembled with the 3D-printed holders and the medium reservoirs in chip holders were filled with cell culture medium (1 ml). (E) The cells were cultured within the assembled chip platforms in 6-well plates for up to one week. (F) Three times per day, impedance measurements were performed and the cell monolayer formation phase was determined optically using bright field microscope images. For each phase, representative fluorescent staining was conducted in addition to evaluate tight junction-associated protein ZO-1 (see section Immunofluorescent staining below).

HUVECs (passage 2-5, PromoCell, #C-12203) were cultured with endothelial cell growth medium (PromoCell, #C-22010) in gelatin-coated (0.1% v $v^{-1}$ in water, STEMCELL, #07903) T25 flasks (Nunc™, #156340). A slightly adapted process was established for the membrane and glass-bottom chips used in static conditions (**Figure 11**). First, the chips were immersed in ethanol (70% v $v^{-1}$ in water) for 20 min, dried under a sterile bench, and coated with fibronectin (50 µg $ml^{-1}$ Corning™, #354008) for 1 h at 37°C. After fibronectin aspiration and complete drying of the chips, 15 µl of medium was added to the medium reservoir containing 15k HUVEC (1000 cells $µl^{-1}$).

#### *4.3.3. Immunofluorescent staining:*

Phosphate-buffered saline (PBS with $Ca^{2+}$ and $Mg^{2+}$, Gibco™, #14040091) was used for all washing steps and dilutions in the following procedure. Cells were washed once and then fixed with 4% formaldehyde (Thermo Scientific™ #28906) for 15 minutes at room temperature (RT). Next, the cells were washed three times, permeabilized with 0.2% Triton-X for 20 minutes, and blocked with 1% normal goat serum for 30 minutes at RT. The samples were then exposed to primary antibody against the tight junction-associated marker ZO-1 (1:400, Cell Signaling, #15652) and the endothelial specific cell junction marker PECAM-1 (1:400, Cell Signaling,

#3528S) for 2 hours at RT. After three washes, samples were treated with the secondary antibodies (1:800, Invitrogen, #A11007 and #A11001) for 1 hour at RT in the dark. Three additional washes followed with Hoechst 33342 (1:5000, Invitrogen™, #H3570) added to the first wash. Fluorescence images were taken using an inverted fluorescence microscope (IX50, Olympus) with cellSens Standard (1.18).

*4.3.4. Analysis of the tight junction barrier formation and cell morphogy:*

To quantify the intensity of the tight junction formation, fluorescence images were analyzed. The cells were segmented using Cellpose 2.0, model type "cyto2" and a set diameter of 100 pixels (see **Figure S1b**).[45] To measure the intensity of ZO-1 and PECAM-1, the mean fluorescence intensity of each cell contour was quantified (see **Figure S1**). For this purpose, the labels for the cell borders were enlarged to a radius of 5 pixels (~1.6 µm) (see **Figure S1c**). For mean cell soma intensity measurements, the masks were eroded to a radius of 10 pixels (~3.2 µm) (see **Figure S1f**). Finally, the ZO-1 and PECAM-1 intensity was normalized (mean cell border intensity/mean cell soma intensity) (see **Figure S1h**) as due to the reflective effect of the gold electrodes, the fluorescence intensity obtained from cells grown on the electrodes was much higher than that of cells grown between the electrodes. Only labels larger than 400 $\mu m^2$ (~twice the size of a nucleus) were included in the analyses. To assess cell morphology and the amount of cell-cell contacts, the same segmentation approach as described above was used. Quantification was performed using pyclesperanto 0.10.3[46].

### 4.4. End-to-end solution for monolayer formation based on ECIS data using Conv1d and KAN neural networks

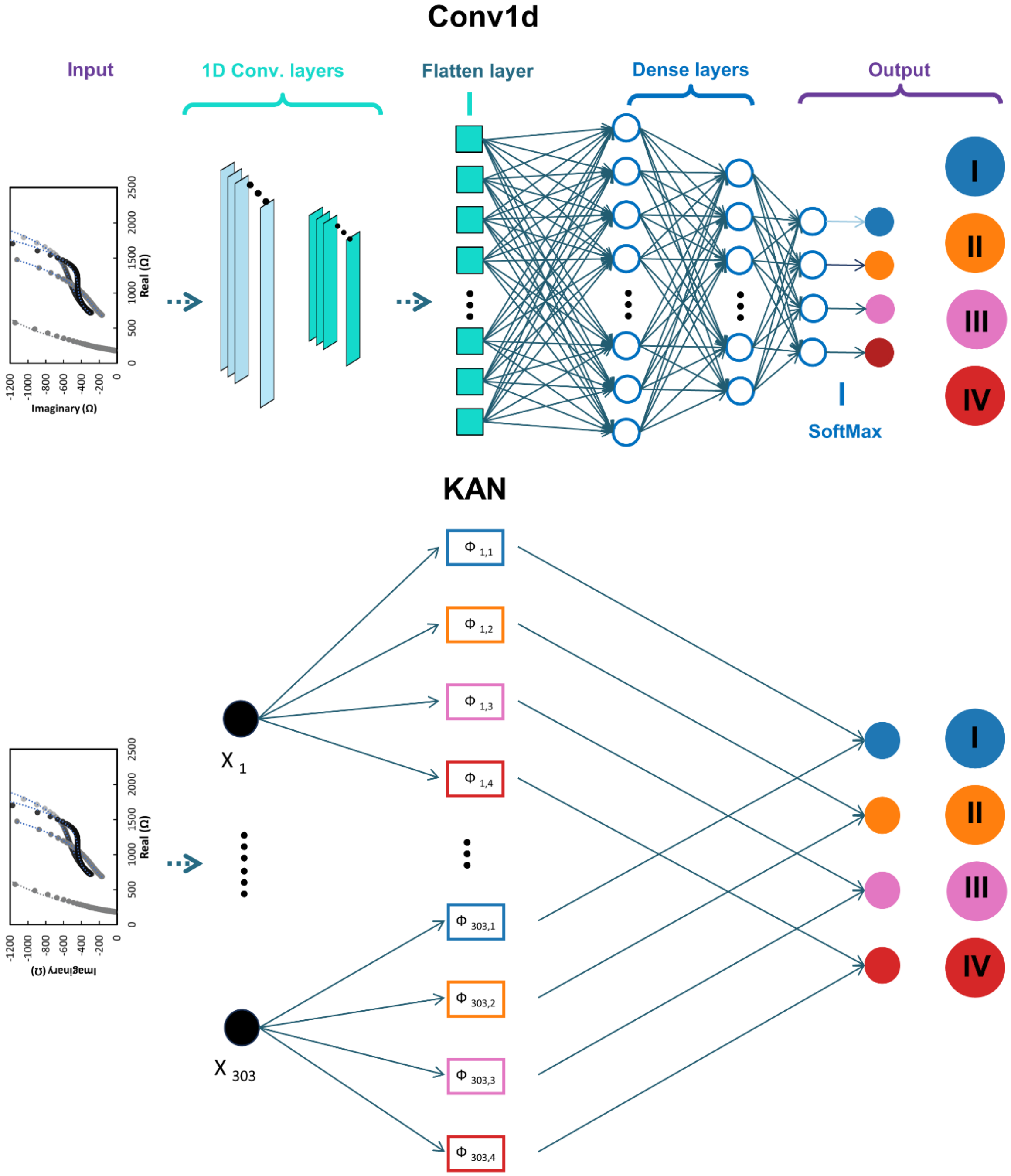

**Figure 12.** Conv1d workflow for classifying EIS impedance spectra into the four phases of cellular monolayer formation. ECIS data were used as input and processed using a 1D convolutional neural network (1D Conv.) for feature extraction. After flattening the 1D convolutional data, two fully connected dense layers and Softmax were used for classification. All processes were implemented in Keras with a TensorFlow backend. KAN workflow for classifying the same dataset, only one layer of activation functions was applied and every input data ($\boldsymbol{X_{input}}$) corresponding to have four different activation functions ($\boldsymbol{\phi}_{input,\ phases}$). The result is shown here under "Output" in the form of four colored circles labelled I, II, III, and IV representing the cellular monolayer formation phases.

To automatically classify ECIS impedance spectra into the four phases of an emerging cellular monolayer, two light neural networks were developed (**Figure 12**). The Conv1d neural network was implemented using Keras with a TensorFlow backend. The proposed neural network model consisted of two 1D convolutional layers (1D Conv.) both with a kernel size of 3 and filters numbers of 120 for automatic feature extraction, followed by two dense layers and a softmax layer for feature classification. A dataset including 303 Nyquist curves collected from the glass-bottom chips annotated with four different labels according to the corresponding cellular growth phase was used to train and test the neural model. Each input instance included two channels, the real and imaginary components of impedance, as shown in the Nyquist diagram. Each channel contains 200 frequency response points. Thus, the input instance of the neural network had 2 channels with a window size of 200. The Rectified Linear Unit (ReLU) function was selected as the activation function of each neuron. KAN networks is more simple compare to the Conv1d networks. Only one KAN-layer was applied and this fully-connected layer, with a 1D function placed on each edge can be regard as a learned activation function. The main difference to KAN is that Conv1d activation functions are on nodes, whereas the activation functions of the KAN layer are on edges. [47,48].

Both models were trained using the categorical cross-entropy loss function and the Adam optimizer with a 0.0002 learning rate for Conv1d and 1 for KAN. KAN models were all trained with grid rang from 0 to 1 and the parameterized activation function in 3 order of pieceweis polynomial. Each model is trained for 100 epochs without an early stop. A five-fold cross-validation method was used to verify the light neural network for the classification task. This dataset was equally divided into five parts, four of which were used as the training set and the remaining one as the test set iteratively. Another dataset with 123 Nyquist curves collected from barrier chips employing ultra-thin nanoporous membranes were also used to test the model trained using data collected from the glass-bottom chip. Accuracy and precision, as described by the ***F1 score*** and macro ***F1 score*** metrics, were used to evaluate the performance of the light-weight neural network model. The accuracy rate was calculated using **Equation 2**. Precision was defined as the ratio of correctly predicted positive observations (i.e., ***TP + TN***) divided by the total predicted positive observations (**Equation 3**). Therefore, precision measures the ‘exactness’ of the model. Recall, also known as model sensitivity, measures the proportion of positives that fwere correctly predicted by the model (**Equation 4**).

$$Accuracy = (TP + TN) / (TP + FP + FN + TN) \quad (2)$$

$$Precision = TP / (TP + FP) \quad (3)$$

$$Recall = TP / (TP + FN) \quad (4)$$

$$F1\ Score = 2 \times (Recall \times Precision) / (Recall + Precision) \quad (5)$$

$$Macro\ F1\ Score = \frac{\sum_{i=1}^{n} F1_i}{\mathrm{n}} \quad (6)$$

where ***TP*** is the number of true positives: The cases in which the model predicted Positive, and the actual class was also Positive. ***TN*** is the number of True Negatives: the model decided Negative, and the true classification (based on microscopic data) resulted also in Negative. ***FP*** is the number of False Positives: the model predicted Positive, but the actual class was Negative. ***FN*** is the number of False Negatives: The model predicted Negative, but the actual class was Positive. n is the number of classes (in this study 4 cell status phases). ***i*** represent a specific class (one of the 4 phases of cell status). The reliable identification of phase four was extremely important. ***F1 score*** is a composite metric that is defined as the harmonic mean of precision and recall that assigns equal weight to both. **Equation 5**, **6** show how ***F1 score*** and ***macro F1 score*** were calculated respectively. The ***macro F1 score*** is a metric for evaluating the performance of a model in multi-class problems. It was computed by first calculating the ***F1 score*** for each class individually, and then taking the average of these ***F1 scores***. This means that when calculating the ***macro F1 score*** each class is given equal importance regardless of the number of samples in each class. The significance of the ***macro F1 score*** is that it assigns equal weight to each class, ensuring that classes with fewer samples are equally important in the overall assessment. The ***macro F1 score*** was particularly useful in cases of class imbalance, as it ensures that the model performs well across all classes, not just the ones with a larger number of samples.[49] All above discussed performance metrics were combined to evaluate the performance of our model according to the cross-system validation.

**Supporting Information**

Supporting Information is available from the Wiley Online Library or from the author.

**Acknowledgements**

The work was partially funded by Niedersächsische Landesregierung in the frame of the coordinated project “Micro Replace Systems”.

Financial support for V.K. was provided by the Volkswagen Stiftung through the funding initiative Change of Course (Kurswechsel).

We acknowledge support by the Open Access Publication Funds of Technische Universität Braunschweig.

Bo Tang and Victor Krajka contributed equally to this work.

**Conflict of Interest**

The authors declare no conflict of interest.

**Data Availability Statement**

The data that support the findings of this study are available from the corresponding author upon reasonable request.

# Supporting Information

## Note 1: Cell analysis procedure

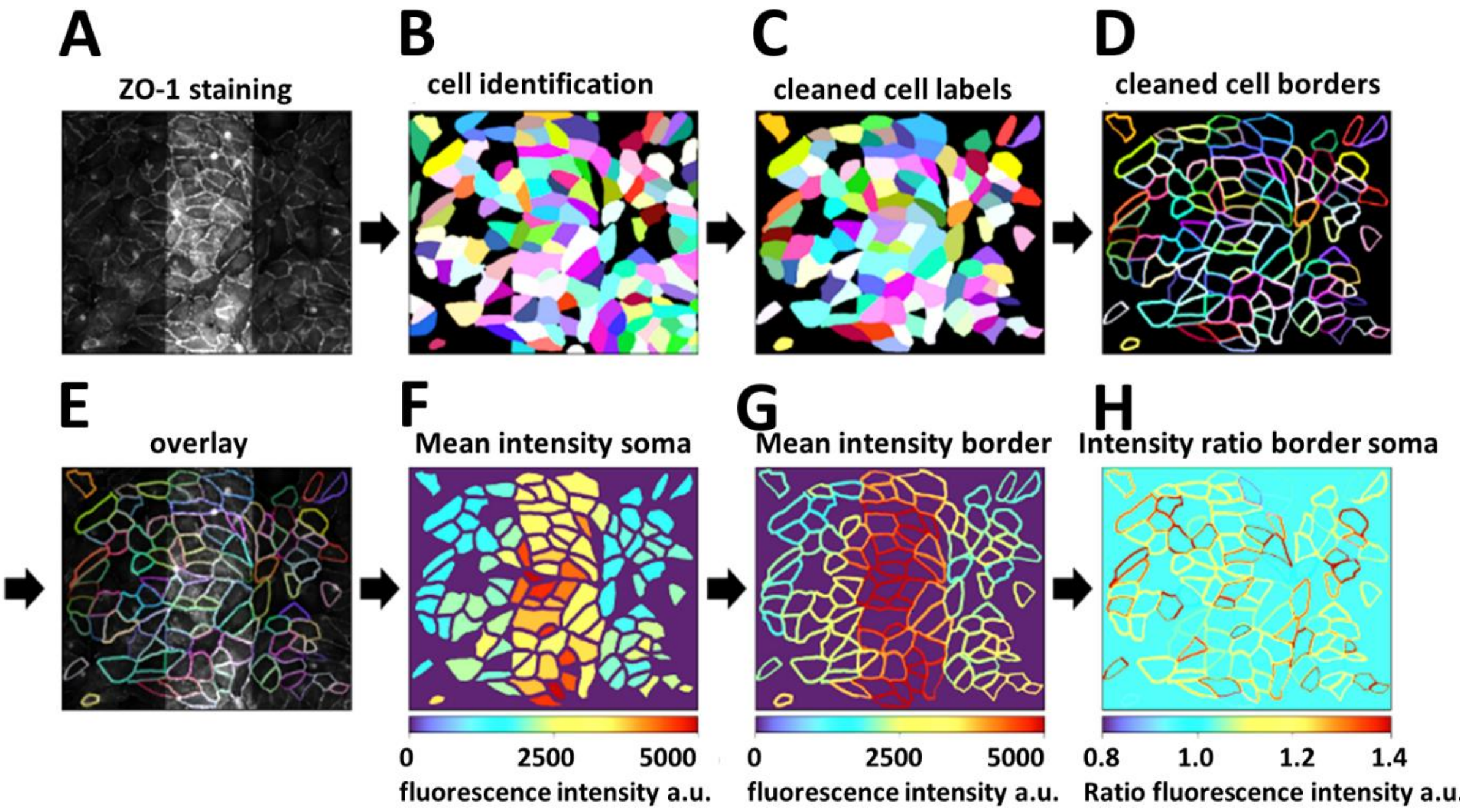

**Figure S1.** Cell analysis procedure using cellpose 2.0. (A-B) First, cells were segmented based on ZO-1 immunofluorescence staining images (size: 1376 x 1038 pixels, scale: 3.1 pixels/µm) using cellpose (model type "cyto2" and set diameter: 100 pixels). (C-D) Cells truncated by image boundaries were excluded. (E) The segmentation was validated by superimposing the cell boundary labels on the original ZO-1 image. (F-G) For ZO-1, fluorescence intensity was measured in the soma and the cell borders. For the average cell soma intensity measurements, the masks were eroded to a radius of 10 pixels (~3.2 µm), and the cell boundary labels were enlarged to a radius of 5 pixels (~1.6 µm). (H) Fluorescence intensity was normalized (ratio of border to soma) to compensate for differences in fluorescence intensity due to the presence of electrodes. Only labels larger than 400 µm$^2$ (~twice the size of a nucleus) were included in the analyses.

**Note 2: Equivalent circuit analysis**

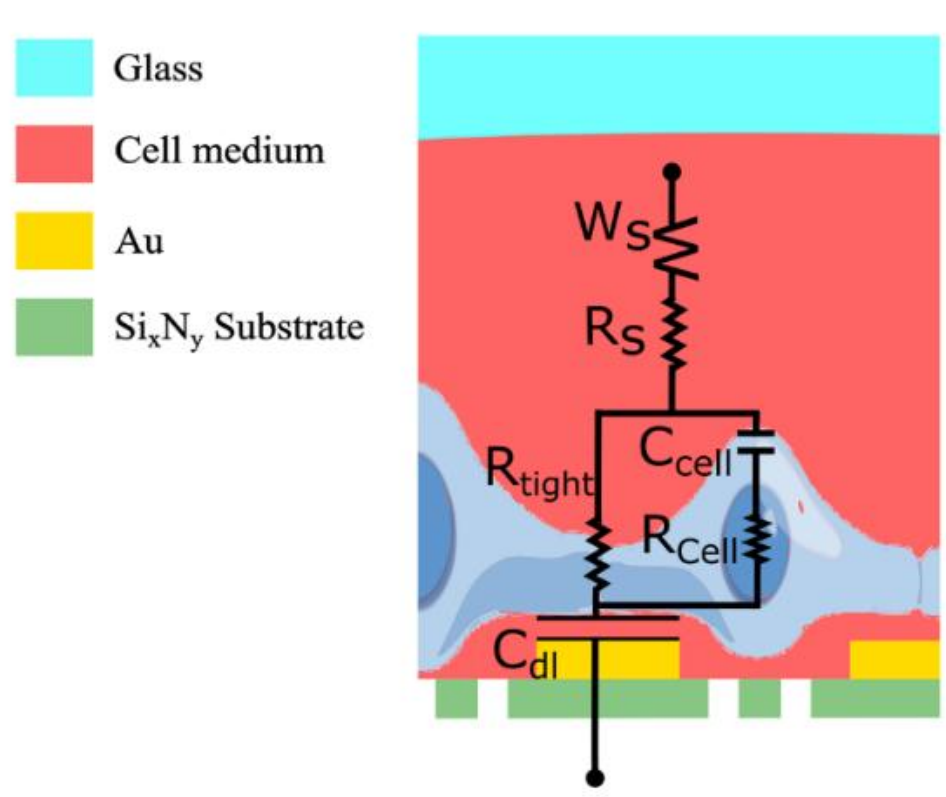

**Figure S2.** Equivalent circuit model showing electrical components of the ECIS circuit. Here, $R_s$ denotes the resistance of the culture medium, which is connected in series with the Warburg element $W_s$ to indicate possible electrochemical reactions in the cell medium. The double-layer capacitance $C_{dl}$ represents the polarization of the microelectrode surfaces. The paracellular route is described by the resistive element $R_{tight}$ representing the tightness of intercellular contacts (tight junctions). The parallel transcellular path includes $C_{cell}$, representing the cell membrane capacitance, and the intracellular resistance $R_{cell}$

**Table S1.** Results of fitting the equivalent circuit given in Figure S2 to the measured Nyquist diagrams as given in **Figure 2** for four phases of the monolayer cell growth on membranes together with respective root mean square error (RMSE), which is the standard deviation of the residuals (not of the measurement values).

| | $R_S$ (Ω) | $W_S$ (Ω) | $W_S$-T | $W_S$-P | $C_{cell}$ (μF) | $R_{sys}$ (Ω) | $C_{sys}$ (μF) | $C_{dl}$ (μF) | $R_{tight}$ (Ω) |
|---|---|---|---|---|---|---|---|---|---|
| **Phase I** | 183 | 3.5 | 3.6×10$^{-06}$ | 0.8 | 3.4×10$^{-08}$ | 698.1 | 2.6×10$^{-07}$ | 2.0×10$^{-07}$ | 51 |
| **RMSE** | 0.2 | 1.0 | 1.7×10$^{-07}$ | 0.033 | 2.6×10$^{-09}$ | 86.3 | 9.2×10$^{-08}$ | 0.1×10$^{-07}$ | 6.5 |
| **Phase II** | 206 | 185.8 | 2.2×10$^{-06}$ | 0.5 | 2.1×10$^{-08}$ | 636.1 | 9.0×10$^{-07}$ | 1.6×10$^{-08}$ | 196 |
| **RMSE** | 1.6 | 6.2 | 5.0×10$^{-08}$ | 0.004 | 2.9×10$^{-10}$ | 5.813 | 1.0×10$^{-08}$ | 4.5×10$^{-10}$ | 2.9 |
| **Phase III** | 210 | 209.8 | 2.2×10$^{-06}$ | 0.6 | 1.2×10$^{-08}$ | 817.6 | 5.9×10$^{-07}$ | 6.0×10$^{-09}$ | 435 |
| **RMSE** | 1.9 | 9.2 | 4.3×10$^{-08}$ | 0.004 | 1.9×10$^{-10}$ | 2.5261 | 6.3×10$^{-09}$ | 1.3×10$^{-10}$ | 4.6 |
| **Phase IV** | 165 | 303.9 | 2.4×10$^{-06}$ | 0.5 | 1.8×10$^{-08}$ | 553.5 | 5.4×10$^{-09}$ | 5.8×10$^{-09}$ | 446 |
| **RMSE** | 3.0 | 16.3 | 6.9×10$^{-08}$ | 0.003 | 4.6×10$^{-10}$ | 4.1067 | 5.4×10$^{-09}$ | 2.3×10$^{-10}$ | 9.2 |

The values of the individual components of the equivalent circuit as obtained in the different phases are given in Table S1. In phase I, $C_{dl}$ assumed a value of 2.0×10$^{-07}$ µF. As the cell density increased, the membrane and electrode surface became gradually more covered, and $C_{dl}$ lowered to 1.6×10$^{-08}$ µF. Once a complete cellular monolayer had formed (phases III and IV), $C_{dl}$ reached the lowest value around 6 ×10$^{-9}$ µF.

## Note 3: Conv1d vs. KAN

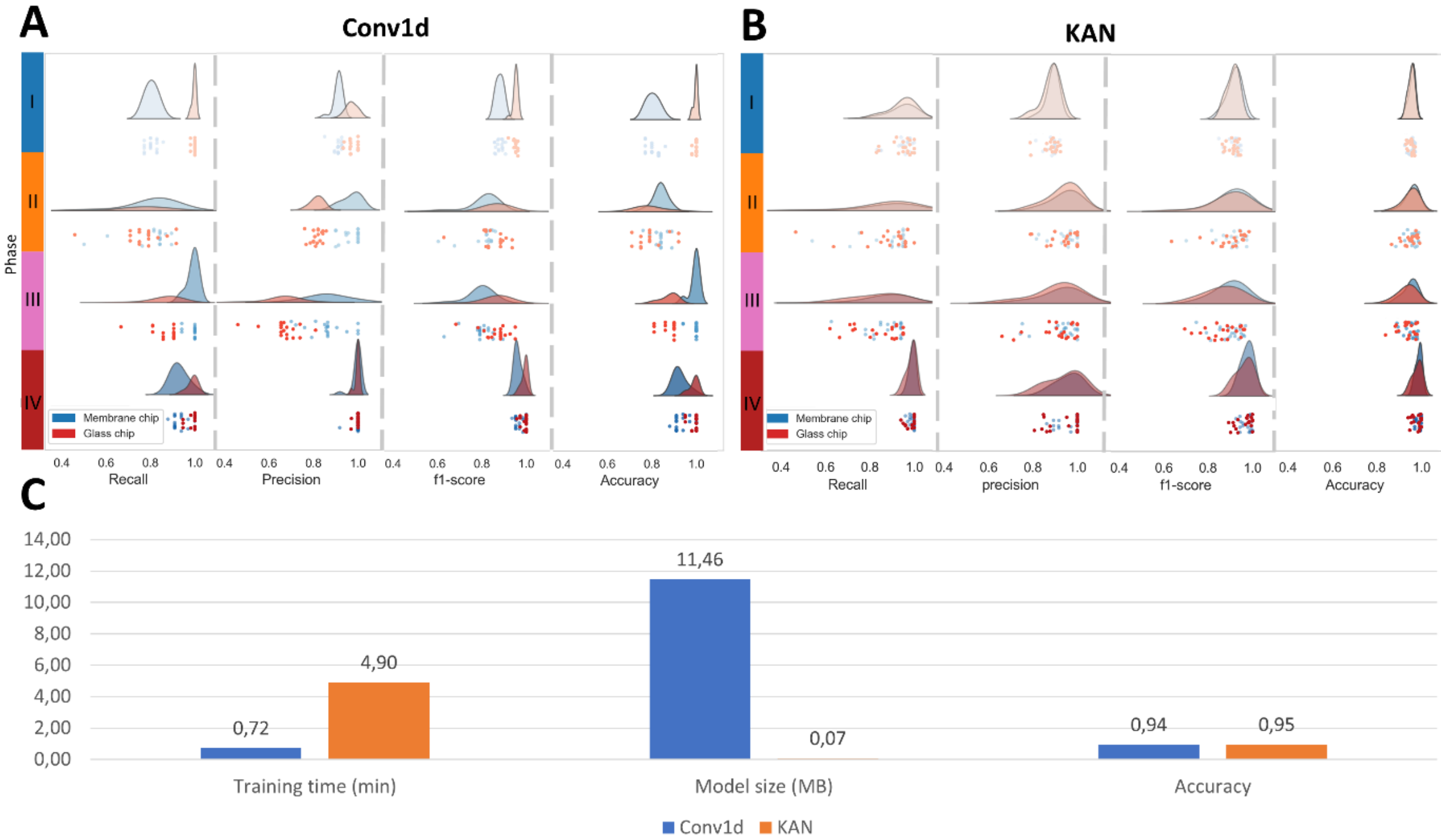

**Figure S3.** (A, B) Rain-cloud plots for all performance metrics for four defined monolayer-cell membrane phases as obtained with Conv1d and KAN networks. (C) Comparison of model efficiency in Conv1d and KAN, in which KAN used longer training time and less data to generate a significantly smaller model but with similar result.

## Note 4: Effect of the barrier modulators

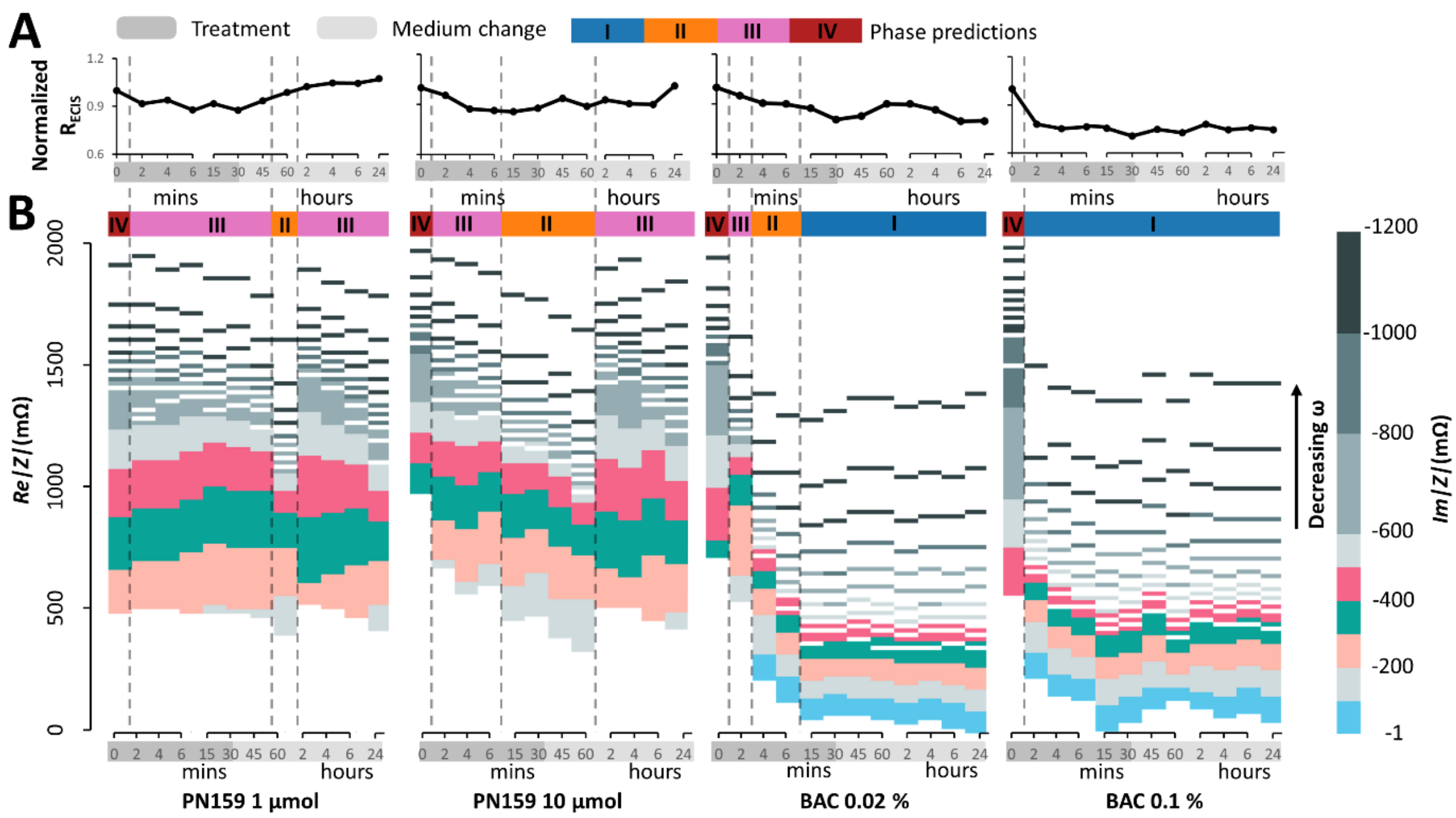

**Figure S4.** Monitoring monolayer integrity during and after drug administration. (A) Normalized $\boldsymbol{R_{ECIS}}$ (Ohmic resistance contribution to the measured total impedance) reveals reversible (PN159) and irreversible (BAC) effects on the cell monolayer. (B) Color-coded Nyquist chronograms based on the $Im|Z|$ values show substance-specific and concentration-dependent impacts on monolayer integrity during and after drug treatment.

**Note 5: Development of ZO-1 and PECAM-1 fluorescence with progressive maturation**

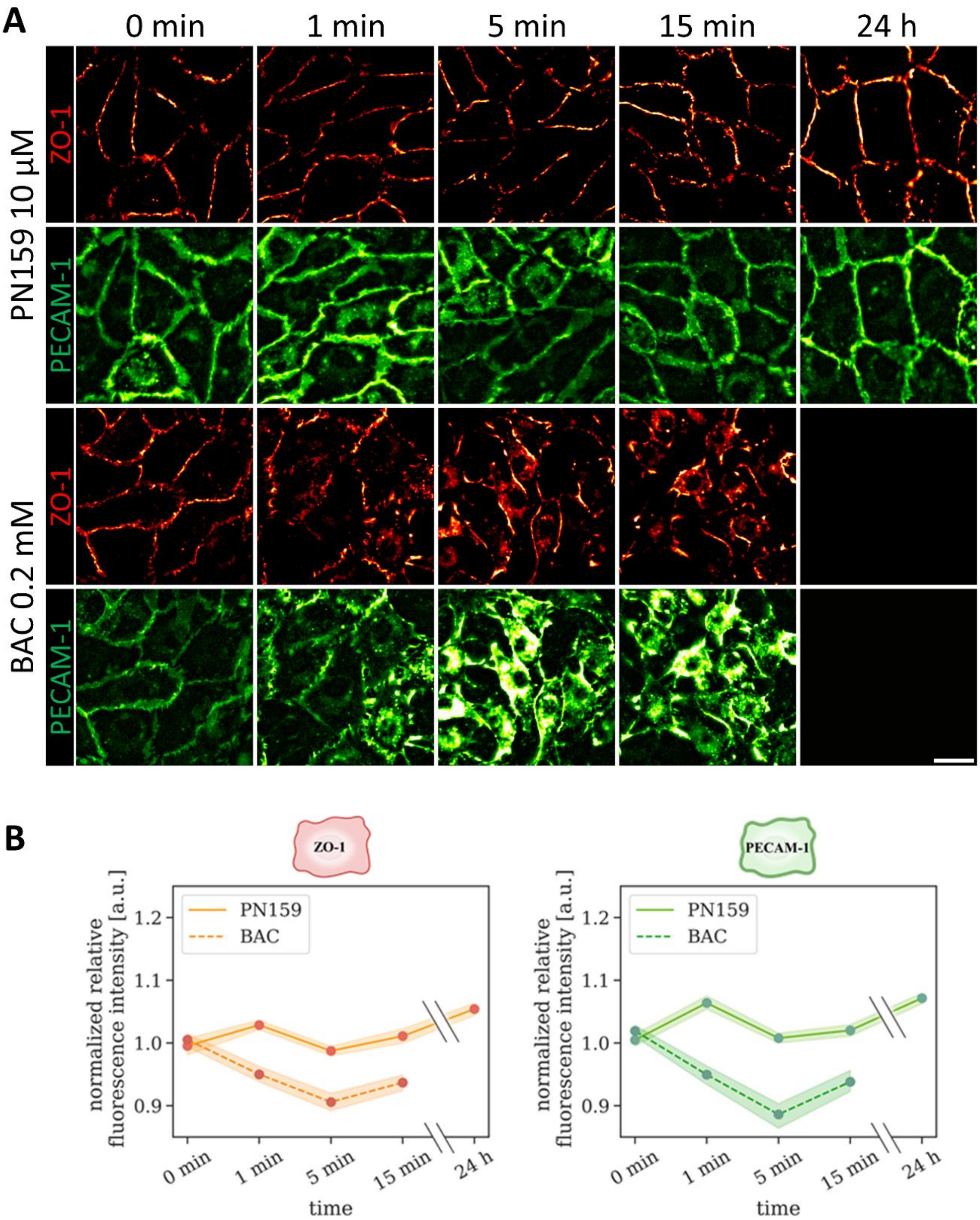

**Figure S5.** Effect of compounds on cell morphology and cell junction-associated proteins over time. (a) Immunofluorescence staining against ZO-1 (red hot) and PECAM-1 (green hot) after treatment with PN159 (10 µM) or BAC (0.02%) at different time points. The compounds were washed out after 30 min. Scale bar represents 10 µm. (b) Evaluation of relative cell outline intensity (intensity ratio: cell outline/soma) normalized to the initial value (at 0 min) shows virtually no or minor changes in ZO-1 and PECAM-1 intensity during and after treatment. The same drug concentrations were used as in (a). Error bands indicate 95% confidence intervals.